\documentclass[11pt,english,a4paper,floatfix]{article}
\usepackage{jcappub}
\usepackage{enumitem}
\pdfoutput=1
\usepackage{bm}
\usepackage{amsfonts}
\usepackage{hyperref}
\usepackage{graphicx}
\usepackage[justification=centering]{subfig}
\usepackage{caption}
\usepackage{relsize,exscale}
\usepackage{array}
\usepackage{ctable}
\usepackage{cellspace}
\newcolumntype{?}{!{\vrule width 0.12em}}

\newcommand{\be}{\begin{equation}}
\newcommand{\ee}{\end{equation}}
\newcommand{\bea}{\begin{eqnarray}}
\newcommand{\eea}{\end{eqnarray}}

\renewcommand{\vec}[1]{\boldsymbol{#1}}

\newcommand{\dv}[2]{\frac{\mathrm{d}#1}{\mathrm{d}#2}}
\newcommand{\pdv}[2]{\frac{\partial#1}{\partial#2}}
\renewcommand{\d}{\mathrm{d}}

\usepackage{lmodern}
\usepackage{slantsc}

\newcommand{\CLASS}{\textsc{class}}
\newcommand{\CAMB}{\textsc{camb}}
\newcommand{\CosmoMC}{\textsc{CosmoMC}}

\begin{document}


\title{Updated constraints on decaying cold dark matter}

\author[a]{Andreas Nygaard,}
\author[a]{Thomas Tram,}
\author[a]{Steen Hannestad}

\affiliation[a]{Department of Physics and Astronomy, Aarhus University,
 DK--8000 Aarhus C, Denmark}
\emailAdd{andreas@phys.au.dk}
\emailAdd{thomas.tram@phys.au.dk}
\emailAdd{steen@phys.au.dk}

\abstract{
In this paper we update the constraints on the simple decaying cold dark matter (DCDM) model with dark radiation (DR) as decay product. We consider two different regimes of the lifetime, i.e. short-lived and long-lived, and use the most recent CMB data from Planck (2018) to infer new constraints on the decay parameters with which we compare the constraints inferred by the previous Planck data (2015). We hereby show that the newest CMB data constrains the fractional amount of DCDM twice as much as the previous data in the long-lived regime, leading to our current best 2$\sigma$ upper bound of $f_{\rm dcdm}<2.44\%$. In the short-lived regime, we get a slightly looser 2$\sigma$ upper bound of $f_{\rm dcdm}<13.1\%$ compared to the previous CMB data. If we include Baryonic Acoustic Oscillations data from BOSS DR-12, the constraints in both the long-lived and the short-lived regimes relax to $f_{\rm dcdm}<2.62\%$ and $f_{\rm dcdm}<1.49\%$, respectively. We also investigate how this model impacts the Hubble and $\sigma_8$ tensions, and we find that each of the decay regimes can slightly relieve a different one of the tensions. The model can thus not accommodate both tensions at once, and the improvements on each are not significant. We furthermore improve on previous work by thoroughly analysing the impacts of short-lived DCDM on the radiation density and deriving a mapping between short-lived DCDM and a correction, $\Delta N_{\rm eff}$, to the effective number of massless neutrino species.
}

\maketitle


\section{Introduction}
Ever since the first inference of the existence of dark matter almost a century ago, the nature of it has remained elusive. 
Significant progress has been made in establishing its properties. For example it is known that dark matter can only interact weakly with standard model particles (perhaps only through gravitation), and that it must be \textit{cold} in the sense of having sufficiently small thermal velocity so that small scale structure can form.
The standard $\Lambda$CDM model of cosmology is based on dark matter with exactly these properties and in general provides an excellent fit to observational data. 

However, despite many advances in both theory and observations much still remains unknown about the nature of dark matter. For example dark matter could have significant interactions, either with itself or with other particles in a dark sector, separate from the standard model.
Such models have been invoked as possible explanations of a variety of anomalies in cosmology, such as the missing satellites problem, the cusp-core problem, and the Hubble tension.

In this paper we investigate the possibility that dark matter is cold and consists of massive particles, but that these particles are unstable and decay.
The decay product cannot simply be electromagnetic radiation, since that would be detectable, so instead it is assumed that the decaying cold dark matter (DCDM) decays into massless particles in a dark sector. In line with previous work on the topic we shall refer to the decay product as \textit{dark radiation} (DR), so that we can sketch the process as 
\begin{align*}
    X_{_{\mathrm{dcdm}}}\xrightarrow{}\gamma_{_{\mathrm{dr}}}\,.
\end{align*}
When comparing a model where all dark matter is DCDM with our observational data (Planck, etc.), we need a very large lifetime which renders our DCDM basically stable. Because of that, we add another degree of freedom to our model which is the initial fraction of DCDM to the total amount of cold dark matter (CDM),
\begin{equation}
    f_{\mathrm{dcdm}}=\frac{\Omega^{\rm ini}_{\mathrm{dcdm}}}{\Omega^{\rm ini}_{\mathrm{dcdm}}+\Omega_{\mathrm{scdm}}}\,,
\end{equation}
where SCDM is the stable CDM component. The physical interpretation of having this fractional decay could be~\cite{Poulin_16}:
\begin{enumerate}[label={\arabic*)}]
    \item CDM is multi-component such that only a part of the dark matter decays, 
    \item all dark matter decays, but the decay product is both dark radiation and a stable CDM component with fraction $(1-f_{\mathrm{dcdm}})$.
\end{enumerate}
Since the amount of dark matter decreases, we define $\Omega^{\mathrm{ini}}_{\mathrm{dcdm}}$ as the density parameter of DCDM today as if none of it had decayed. We otherwise adopt the same notation as in Ref.~\cite{MaBert}, which is now standard.

\subsection{Previous work and constraints}\label{sec:previous}
The field of decaying cold dark matter has been growing ever since the first analysis by Ref.~\cite{Ichiki_2004} in 2004, where the simple model with dark matter decaying into dark radiation was also used. A lot has happened in the last two decades with the field being much wider now with numerous models of varying sophistication. The different regimes of the decay rate, $\Gamma_{\mathrm{dcdm}}$, is investigated in Ref.~\cite{Poulin_16} and the fractional amount of DCDM is constrained to $f_{\mathrm{dcdm}}<3.8\times10^{-2}$ using Cosmic Microwave Background (CMB) data from Planck-2015. This is in agreement with the results found in Ref.~\cite{Chudaykin_2016} where the same CMB data was used. In this paper we update the results and constraints from Ref.~\cite{Poulin_16} using the newest Planck-2018 data as well as Baryonic Acoustic Oscillations (BAO) data. 

It is well known that allowing dark matter to decay can potentially relieve the Hubble tension as proposed in Ref.~\cite{Berezhiani_2015}, and this has been demonstrated numerous times in the literature. This includes Ref.~\cite{Pandey_2020} where both the Hubble tension and the milder $S_8$ tension of matter fluctuations between $\Lambda$CDM-model and the model-independent measurements in the local Universe were relieved using CMB data from Planck-2018 and the same DCDM model as we are using in this paper. Including other data sets such as BAO data and intermediate-redshift data, however, leads to only a slight decrease in both tensions. Another attempt at relieving the Hubble tension is made in Ref.~\cite{Chudaykin_2018} where the tension is reduced by $1.5\sigma$ using Planck-2015 data, BAO data, Redshift Space Distortion (RSD) measurements and a DCDM model with decay time after recombination. By allowing the CMB lensing power amplitude to be a free fitting parameter, the tension is reduced by $3.3\sigma$. They furthermore found an upper limit on the fractional amount of DCDM at the level $f_{\rm dcdm}\lesssim5\%$ for DCDM decaying after recombination. The fractional amount of short-lived DCDM is also constrained in Ref.~\cite{Xiao_2020} where CMB data from Planck-2015, BAO data, and RSD measurements are used to infer the upper limit $f_{\rm dcdm}\lesssim 2.73\%$, but this only leads to a slight reduction of the Hubble tension. They furthermore show the impacts of DCDM on the kinetic Sunyaev-Zel'dovich effect and note that this could lead to further constraints on DCDM in the future.

Better and more data has allowed the constraints to be more refined, with different studies using different combinations of data sets. An upper bound on the decay rate of late time DCDM at $\Gamma_{\mathrm{dcdm}} \leq (175 \,\mathrm{Gyr})^{-1}$ is inferred in Ref.~\cite{Enqvist_2020} by using CMB data and cluster counts from Planck-2015 along with weak lensing data from KiDS450, while a more tight constraint of $\Gamma_{\rm eff}<9.1\times 10^{-9}\,{\rm Gyr^{-1}}$ is found in Ref.~\cite{Oldengott_2016} to the effective decay rate by using Planck-2015 data and analysing cosmic reionisation and dark matter decay simultaneously. A different approach is taken in Ref.~\cite{Blanco_2019} where they assume that the decay product is detectable, i.e. a pair production of an elementary particle from the Standard Model (quarks, leptons or bosons), which leads to lower limits on the lifetime in the range $\tau\sim(1-5)\times 10^{28}\,\mathrm{s}$ using data from the isotropic gamma-ray background.

There are many studies beyond the simple DCDM model, where the decay product is only DR, and a more agnostic approach is found in Ref.~\cite{Bringmann_2018} where the conversion of dark matter to dark radiation is treated in a more general aspect without assuming the transition being caused by a decay. They then find the impacts of the conversion on the CMB spectrum as well as constraints on their general model parameters using \CAMB\, and \CosmoMC, which lead to a reduction in the Hubble tension. They then treat a conversion model in detail where dark matter particles interact via a light mediator particle leading to Sommerfeld-enhanced self-annihilation.

Another, more general analysis is done in Ref.~\cite{Dienes_2017} where they investigate Dynamical Dark Matter in which the dark sector is comprised of a large ensemble of particles with different decay widths. Their analysis shows that the constraints allow energy scales ranging from GeV scale to the Planck scale, but the cosmological abundance of the dark sector today must be spread across an increasing number of states in the ensemble as the energy scale is decreased down to GeV scale from the Planck scale. A third continuation of the DCDM model is found in Ref.~\cite{Raveri_2017} which features Partially Acoustic Dark Matter, where a subdominant part of the dark matter is strongly coupled to the DR fluid, and this combined fluid undergoes acoustic oscillations below the effective dark sound horizon. This is used to reduce the tensions in the Hubble constant and the lensing amplitude between CMB data and direct measurements. They find that even though the model can be used to reduce the tensions separately, it cannot accommodate both at once, since additional CDM is required by the CMB data to preserve the shape of the acoustic peaks.

It has recently been increasingly popular to assume a slightly warm component of the decaying dark matter sector, and an analysis of the background equations can be found in Ref.~\cite{Vattis_2019}, where the cold dark matter decays to both DR and a warm daughter particle. They show that this is can potentially relieve the Hubble tension, which makes the model very interesting to future work in the field of cosmology. More thorough analyses of the two-body decay model are found in Refs.~\cite{Blackadder:2014wpa,Clark:2020miy,Haridasu:2020xaa}, and the same model has also been shown to potentially relieve the $S_8$ tension in various studies, including Refs.~\cite{Wang:2012eka,Abellan:2020pmw,Abellan:2021bpx}. This is due to the recoil velocity of the massive daughter particle inducing a free-streaming suppression of matter fluctuations. The model has also shown promise when it comes to the problems of small-scale structure, and treatments of this using N-body simulations can be found in Refs.~\cite{Borzumati:2008zz,Peter:2010jy,Peter:2010sz,Wang:2014ina}. More general studies, where both daughter particles can have arbitrary masses, are treated in Refs.~\cite{Aoyama:2011ba,Aoyama:2014tga}. A slightly different model, where the decaying component is warm and the decay product is only DR, is treated thoroughly in Ref.~\cite{Blinov_2020}. The idea of this model is to not modify the evolution of the gravitational potentials, which otherwise leads to inconsistencies with data as in the case of decaying cold dark matter. They find that this can significantly reduce the Hubble tension as well.

\subsection{Outline of this paper}
In this paper, we are updating the constraints on the DCDM model parameters from Ref.~\cite{Poulin_16} using the Markov Chain Monte Carlo (MCMC) sampler \textsc{MontePython}~\cite{montepython} and the Einstein--Boltzmann code \CLASS~\cite{class}. To describe the cosmological framework we will use the following set of parameters:
\begin{equation}
    \Theta = \{\Omega_{\rm b} h^2, \Omega_{\rm cdm} h^2, h^2, A_s, n_s, \tau_{\rm reio} \},
\end{equation}
in addition to the decay parameter vector $\{ f_{\rm dcdm}, \Gamma_{\rm dcdm} \}$.
As was also done in Ref.~\cite{Poulin_16}, we will split our MCMC runs into two categories, named \textit{short-lived} and \textit{long-lived}. The reason for this is that the likelihood contour in the full parameter space has a shape which makes convergence exceedingly slow. When the lifetime becomes very short, essentially all DCDM has decayed well before matter-radiation equality. This makes it impossible to constrain $\Gamma_{\rm dcdm}$, and the only constrainable quantities are then $f_{\rm dcdm}$ and the product $\Gamma f_{\rm dcdm}$. This leads to a funnel-like likelihood surface stretching towards very large values of $\Gamma_{\rm dcdm}$. This particular part of the parameter space is best probed using a logarithmic prior on $\Gamma_{\rm dcdm}$, whereas for the long-lived regime, where only a fraction of the dark matter has decayed before the present, a prior which is flat in $\Gamma_{\rm dcdm}$ is more suitable. We will elaborate on the technicalities of this split in section~\ref{sec:MCMC}.

We will furthermore look at an analogous scheme to the short-lived DCDM, i.e. a model with an increase in $N_{\rm eff}$ instead of a decaying dark matter component. This is treated both analytically and numerically using the \CLASS\, code.

The structure of this paper is as follows. We introduce the formal theoretical framework of DCDM in the synchronous gauge in section~\ref{sec:boltzmann}, where both the background equations and the perturbation equations are presented. In section~\ref{sec:Neff_mapping} we will treat the mapping between a correction to $N_{\rm eff}$ and short-lived DCDM by deriving an analytical expression for $\Delta N_{\rm eff}$ corresponding to a short-lived DCDM component and comparing this to numerical results from \CLASS. In section~\ref{sec:MCMC} we will present our results from the MCMC sampler \textsc{MontePython} along with improved constraints in the long-lived and short-lived regimes, and in section~\ref{sec:tensions} we will investigate the impact of our model on the Hubble and $\sigma_8$ tensions. Lastly, we will conclude and summarise in section~\ref{sec:conclusion}.

\subsection{Cosmological data}
As our data sets we use the newest CMB data from Planck-2018~\cite{Planck2018} which has a higher quality in the polarisation data than its predecessor from 2015~\cite{Planck2015}. In all of our computations using Planck-2018, we use both polarisation and temperature likelihoods for both high-$\ell$ and low-$\ell$ as well as the lensing likelihood. When comparing to Planck-2015 data, we of course use the corresponding likelihoods from this data release, which is the same combination used in Ref.~\cite{Poulin_16}.

We furthermore use the BAO data from BOSS data release 12 (DR-12)~\cite{bao}. While the BAO data presumably has little impact in parameter constraints in the short-lived regime, due to the decay happening much earlier than the formation of the BAO, it is quite important in the limit of very long-lived DCDM.

\section{Boltzmann equations for DCDM and DR}\label{sec:boltzmann}
The behaviour of DCDM and DR can be calculated using the Boltzmann equation, which takes the following form for DCDM~\cite{Poulin_16}:

\begin{equation}\label{eq:boltzmann}
    \dv{f}{\tau}=\pdv{f}{\tau}+\pdv{f}{x^i}\dv{x^i}{\tau}+\pdv{f}{p}\dv{p}{\tau}+\pdv{f}{\hat{p}^i}\dv{\hat{p}^i}{\tau}=\pm a\Gamma_{\mathrm{dcdm}}f_{\mathrm{dcdm}}\,,
\end{equation}
with $-$ and $+$ for DCDM and DR respectively.

\subsection{Background equations}
The zeroth moment of the Boltzmann equation, eq.~\eqref{eq:boltzmann}, i.e. integrating it over phase-space and keeping only terms of zeroth order, leads to the continuity equation, which is different for DCDM and DR than for the homogeneous universe in having source terms dependent of the decay rate $\Gamma_{\mathrm{dcdm}}$ with respect to proper time~\cite{Poulin_16}:
\begin{equation}\label{eq:dcdm_fluid}
\begin{aligned}
    \rho'_{\mathrm{dcdm}}&=-3\frac{a'}{a}\rho_{\mathrm{dcdm}}-a\Gamma_{\mathrm{dcdm}}\rho_{\mathrm{dcdm}}\,,\\
    \rho'_{\mathrm{dr}}&=-4\frac{a'}{a}\rho_{\mathrm{dr}}+a\Gamma_{\mathrm{dcdm}}\rho_{\mathrm{dcdm}}\,,
\end{aligned}
\end{equation}
where the prime denotes derivatives with respect to conformal time, $\tau$.

\subsection{Perturbation equations}
One way of obtaining the relevant perturbation equations is again through the Boltzmann equation, eq.~\eqref{eq:boltzmann}. We are doing our calculations in the comoving synchronous gauge, so we need to express the perturbation equations in this gauge. Perturbations in the synchronous gauge can be expressed in the following way using the space-time interval~\cite{MaBert}:
\begin{equation}
    \d s^2=\,a^2(\tau)\left(-\d\tau^2+\left[\delta_{ij}+h_{ij}(\vec{x},\tau)\right]\d x^i\d x^j\right)\,,
\end{equation}
where the scalar part of the perturbation $h_{ij}(\vec{x},\tau)$ can be expressed through its Fourier transform
\begin{equation}
h_{ij}(\vec{x},\tau)=\mathop{\mathlarger{\mathlarger{\mathlarger{\int}}}}\d^3k\,\mathrm{e}^{i\vec{k}\cdot\vec{x}}\left(h(\vec{k},\tau)\hat{k}_i\hat{k}_j +6\eta(\vec{k},\tau)\left[\hat{k}_i\hat{k}_j-\frac{1}{3}\delta_{ij}\right]\right)\,,
\end{equation}
with $h(\vec{k},\tau)$ and $\eta(\vec{k},\tau)$ being the metric perturbations in the synchronous gauge. Integrating the Boltzmann equation, eq.~\eqref{eq:boltzmann}, over phase-space and keeping the first order terms leads to an expression for the evolution of the density perturbation, $\delta_{\mathrm{dcdm}}=\rho_{\mathrm{dcdm}}/\bar{\rho}_{\mathrm{dcdm}}-1$, where the bar denotes the average value as in a homogeneous universe. We can furthermore find the evolution of the divergence, $\theta_{\mathrm{dcdm}}$, of the fluid velocity by taking the divergence of the first moment of eq.~\eqref{eq:boltzmann}, which is found by multiplying the equation by $\overrightarrow{p}/E$ (with $\overrightarrow{p}$ and $E$ being the 3-momentum and the energy of a particle, respectively) and again integrating over phase-space. The resulting equations are
\begin{align}
    \delta'_{\mathrm{dcdm}}&=-\frac{h'}{2}\,,\\
    \theta'_{\mathrm{dcdm}}&=-\mathcal{H}\theta_{\mathrm{dcdm}}=0\,.
\end{align}
These two equations are enough to describe the evolution of DCDM since it per definition is cold, which means that we have neglected all second (or higher) order terms of the momentum in our calculations~\cite{dodelson}. All higher moments would therefore be zero.

The equations turn out a bit more complicated for DR, since this species is not cold and we therefore cannot neglect higher orders of momentum. Following the same procedure as for DCDM in the synchronous gauge, we get the following equations for the evolution of the density perturbations and the velocity divergence:
\begin{align}
    \delta'_{\mathrm{dr}}&=-\frac{2}{3}h'+a\Gamma_{\mathrm{dcdm}}\frac{\rho_{\mathrm{dcdm}}}{\rho_{\mathrm{dr}}}(\delta_{\mathrm{dcdm}}-\delta_{\mathrm{dr}})\,,\\
    \theta'_{\mathrm{dr}}&=\frac{k^2}{4}\delta_{\mathrm{dr}}-k^2\sigma_{\mathrm{dr}}-a\Gamma_{\mathrm{dcdm}}\frac{\rho_{\mathrm{dcdm}}}{\rho_{\mathrm{dr}}}\theta_{\mathrm{dr}}\,.
\end{align}
These certainly do not look as simple as those for DCDM, and we notice the parameter $\sigma_{\mathrm{dr}}$ in the bottom equation which is the next moment of the Boltzmann equation - the shear stress. Generally the evolution of a moment, $l$, will depend on the next moment, $l+1$, which leads to an infinite Boltzmann hierarchy for the moments containing the same information as the momentum-dependent Boltzmann equation itself. We still need to truncate the hierarchy at some large $l$-value to work with it numerically. In the \CLASS\, code, the cut-off $l$-value can be user-specified and is $l_{\mathrm{max}}=17$ by default.

\section{Mapping short-lived DCDM to $N_{\rm eff}$}\label{sec:Neff_mapping}

In the very short-lived regime, all DCDM has decayed well before matter-radiation equality. In this regime the primary effect of DCDM should be to enhance $N_{\rm eff}$ through the decay product, DR. 
We define the correction, $\Delta N_{\rm eff}$, as the ratio between the energy densities of the additional radiation (which is DR) and a single massless neutrino. However, we need to evaluate this ratio after the DCDM has fully decayed, where no more DR is produced and the energy density scales as $a^{-4}$ (like any type of radiation), 
\begin{equation}\label{eq:dN_eff_def}
\Delta N_{\rm eff} = \left.\frac{\rho_{\rm dr}}{\rho^{N=1}_{\nu}}\right|_{t\gg t_d}\,,
\end{equation}
where $\rho^{N=1}_{\nu}=7/8\,(4/11)^{4/3}\rho_{\gamma}$ represents only a single massless neutrino~\cite{Lesgourgues:2018oca}, and $t_d$ is the time of the decay.

We can estimate how $\Delta N_{\rm eff}$ should scale by assuming an instant decay, so the energy density of DR just after the decay equals that of DCDM just before the decay. We can thus evaluate the energy density of DCDM at the time of decay instead of that of DR. We can then scale that to the current time, introducing $a_d$ as the scale factor at the time of decay
\begin{equation}
\Delta N_{\rm eff} \approx \left.\frac{\rho_{\rm dcdm}}{\rho^{N=1}_{\nu}}\right|_{t=t_d} \sim a_d\cdot \frac{f_{\rm dcdm}}{1-f_{\rm dcdm}} \cdot\frac{\Omega_{\rm dm,0}}{\Omega^{N=1}_{\nu,0}}\,.
\label{eq:neff}
\end{equation}
Of course this is a very simplistic calculation, but it shows that, as $a_d \to 0$ (and thus $\Gamma_{\rm dcdm} \to \infty$) the effect of DCDM vanishes and the model becomes identical to standard $\Lambda$CDM.

\subsection{Analytic solution of Boltzmann equations}\label{sec:Neff_analytical}

We can derive an analytic approximation to the Boltzmann equations fairly easily.
First, we define
\begin{equation}
Y = a^3\,\rho_{\rm dcdm}/\rho_{\nu,0}\,, \qquad X = a^4\,\rho_{\rm dr}/\rho_{\nu,0}\,,
\end{equation}
i.e. $X$ and $Y$ are constant in the absence of decays. Here $\rho_{\nu,0}$ is the energy density of all neutrinos today as described by the effective number $N_{\rm eff}$. The Boltzmann equations~\eqref{eq:dcdm_fluid}, can then be expressed in proper time as
\begin{equation}\label{eq:N_eff_boltzmann}
    \dv{Y}{t} = -\Gamma Y(t)\,, \qquad    \dv{X}{t} = a\Gamma Y(t)\,,
\end{equation}
where we have dropped the subscript on the decay rate $\Gamma$. The Boltzmann equation for the decaying component then has the solution
\begin{equation}
    Y(t) =Y_i e^{-\Gamma t}\,,
\end{equation}
\begin{equation}\label{eq:Yi}
    Y_i = a_i^3\frac{(\rho_{\rm dcdm})_i}{\rho_{\nu,0}}=\frac{\Omega^{\rm ini}_{\rm dcdm}}{\Omega_{\nu,0}}=\frac{\Omega_{\rm dm,0}}{\Omega_{\nu,0}}\cdot\frac{f_{\rm dcdm}}{1-f_{\rm dcdm}}\,,
\end{equation}
where the subscript $i$ represents some initial starting point before the decay, and the last equal sign assumes that all DCDM has decayed today, thus making $\Omega_{\rm dm,0}$ both the current density parameter of all dark matter and that of only stable dark matter, since no decaying component is left.

The Boltzmann equation for DR can now be solved using the solution to that of DCDM. We switch coordinates to the scale factor divided by the initial scale factor at the starting point, $\tilde a \equiv a/a_i$, and assume that the Universe is radiation dominated throughout the decay so that $a \propto t^{1/2}$. We furthermore define 
\begin{equation}
    \alpha \equiv \frac{H_i}{\Gamma} = \left(\frac{8\pi G}{3}\rho_{\rm r,0}\,a_i^{-4}\right)^{1/2}\cdot\frac{1}{\Gamma}\,,
\end{equation}
where $\alpha \gg 1$. We then integrate the differential equation from the starting point $\tilde a = 1$ to find the solution for $X$ (a similar result can be found in Ref.~\cite{Scherrer:1984fd})
\begin{equation}
    X(\tilde a) = X_i + \left[\sqrt{\frac{\pi}{2}}{\rm erf}\left(     \frac{\tilde a}{\sqrt{2 \alpha}}  \right)- \frac{\tilde a e^{-\tilde a^2/(2\alpha)}}{\sqrt{\alpha}}\right]\left(\frac{8 \pi G }{3}\rho_{\rm r,0}\right)^{1/4} \frac{\Omega_{\rm dm,0}}{\Omega_{\nu,0}}\cdot\frac{f_{\rm dcdm}}{1-f_{\rm dcdm}}\Gamma^{-1/2}\,,
\end{equation}
where $X_i$ is the initial value of $X(\tilde a)$ proportional to the initial DR component (here we have $X_i=0$ since there is no pre-existing DR component). In the asymptotic limit ($\tilde a \to \infty$) we find
\begin{equation}
    X(\tilde a \to \infty) = X_i + \sqrt{\frac{\pi}{2}}\left(\frac{8 \pi G \rho_{\rm r,0}}{3}\right)^{1/4} \frac{\Omega_{\rm dm,0}}{\Omega_{\nu,0}} \cdot\frac{f_{\rm dcdm}}{1-f_{\rm dcdm}}\Gamma^{-1/2}\,,
    \label{eq:neff2}
\end{equation}
and we finally note that $\Delta N_{\rm eff} = N_{\rm eff} X(\tilde a \to \infty)$ because of our definition of $X$.
Here we notice the scaling relation $\Delta N_{\rm eff} \propto \Gamma^{-1/2} f_{\rm dcdm}/(1-f_{\rm dcdm})$. Inserting numbers in eq.~\eqref{eq:neff2}, we arrive at an approximate relation of the form
\begin{equation}
    \Delta N_{\rm eff} \sim 5.74 \, \Omega_{\rm dm,0} \, h^2 \, \frac{f_{\rm dcdm}}{1-f_{\rm dcdm}}\Gamma_6^{-1/2}\,, 
\end{equation}
where $\Gamma_6 = \Gamma/(10^6 \, {\rm Gyr}^{-1})$.

\subsection{Numerical analysis using \textsc{CLASS}}\label{sec:N_eff_class}

\begin{figure}[t]
    \centering
    \includegraphics{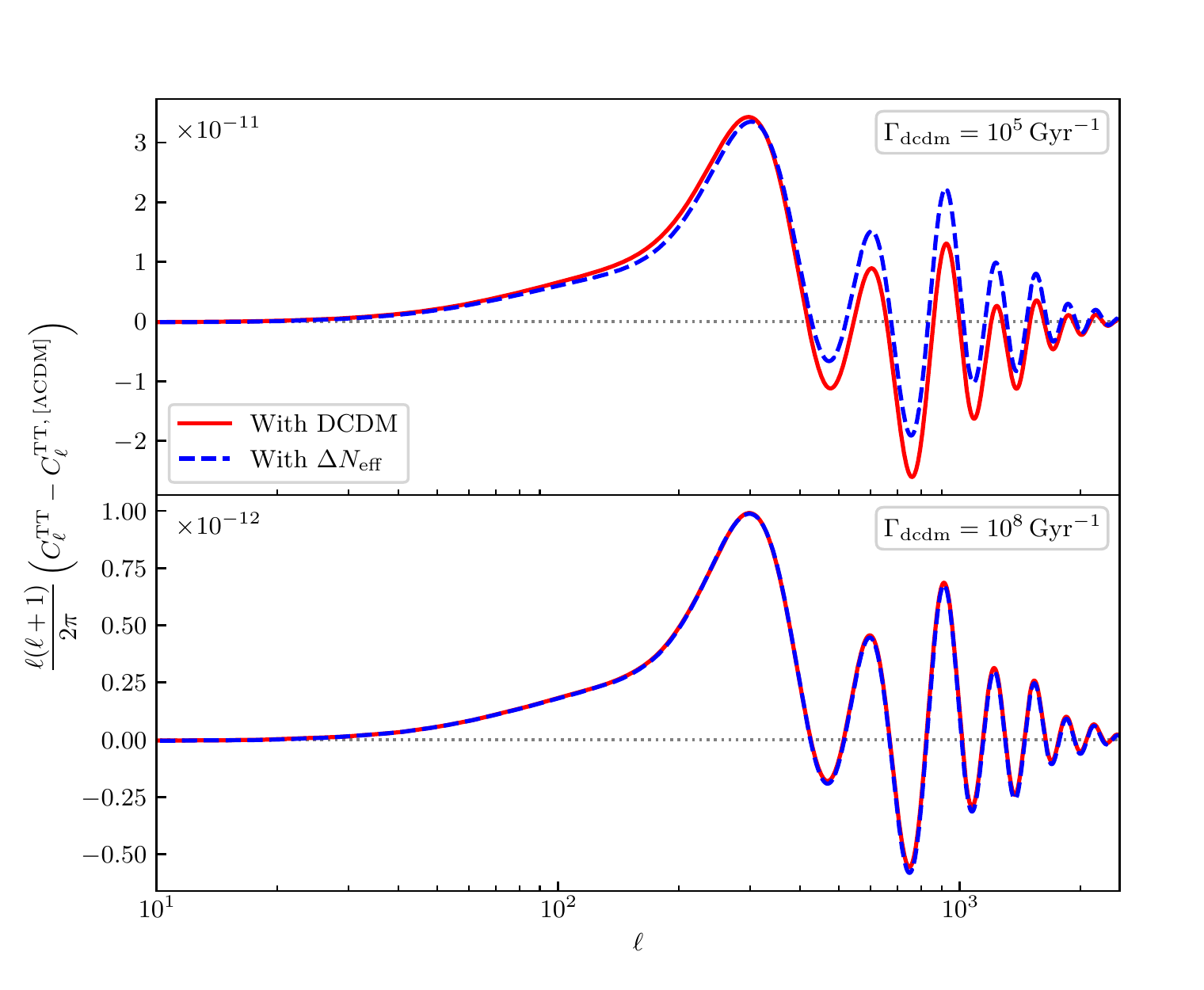}
    \caption{CMB TT power spectra for different values of $\Gamma_{\rm dcdm}$ and the corresponding $\Delta N^{\rm numerical}_{\rm eff}$ (calculated used eq.~\eqref{eq:dN_eff_class}) relative to the power spectrum of the $\Lambda$CDM model with no correction to $N_{\rm eff}$. The additional parameters used for the simulations are found in table~\ref{tab:class_input}, and the fractional amount of DCDM is fixed to $f_{\rm dcdm}=0.2$.}
    \label{fig:CMB}
\end{figure}

Using the \CLASS\, code, we can get an exact numerical correction to $N_{\rm eff}$ caused by short-lived DCDM. The values of input parameters which are held constant in all of the calculations can be seen in table~\ref{tab:class_input} ($N_{\rm eff}$ is modified by $\Delta N_{\rm eff}$ in figure~\ref{fig:CMB}).

\begin{table}[t]
\centering
\setlength\extrarowheight{4pt}
\begin{tabular}{| c c c c c c c  |}
\hline
\multicolumn{1}{| c }{$\Omega_{\rm b}h^2$}&\multicolumn{1}{ c }{$\Omega_{\rm cdm}h^2$}&\multicolumn{1}{ c
}{$h^2$}&\multicolumn{1}{ c
}{$A_s \times 10^{9}$}&\multicolumn{1}{ c
}{$n_s$}&\multicolumn{1}{ c
}{$\tau_{\rm reio}$}&\multicolumn{1}{ c |}{$N_{\rm eff}$}\\
\hline
0.022032  & 
0.11933  & 
0.67556  &
2.215  &
0.9619  &
0.0925  &
3.046 \\
\hline
\end{tabular}
\caption{Table of input parameters in \CLASS\, which are held constant through all calculations in section~\ref{sec:N_eff_class}. $N_{\rm eff}$ is, however, modified with $\Delta N_{\rm eff}$ in figure~\ref{fig:CMB}.}
\label{tab:class_input}
\end{table}

Running \CLASS\, with a high $\Gamma_{\rm dcdm}$ ($>10^3\,\mathrm{Gyr}^{-1}$) ensures that all DCDM has decayed well before today, which means that the energy density of the resulting DR scales like normal radiation. DR and neutrinos therefore scale similarly, which makes the ratio of their energy densities constant after the decay. This ratio is used to define the correction $\Delta N_{\rm eff}$ cf. eq.~\eqref{eq:dN_eff_def}. However, because we do not only have a single neutrino, we need to divide the neutrino density with $N_{\rm eff}$. The numerical value of $\Delta N_{\rm eff}$ can in fact be evaluated at any time after the decay as long as the energy density of DCDM is no longer of any significant size. We, however, evaluate at the current time in order to get a density of DCDM as low as possible,
\begin{equation}\label{eq:dN_eff_class}
\Delta N^{\rm (numerical)}_{\rm eff} = \left. N_{\rm eff}\frac{\rho_{\rm dr}}{\rho_{\nu}}\right|_{t=t_0},
\end{equation}
where $\rho_{\nu}$ is the total energy density of all massless neutrino species.

\begin{figure}[t]
\centering
\subfloat[\label{fig:dNeff_Gamma}]{\hspace{.57\textwidth}}
\subfloat[\label{fig:dNeff_f}]{\hspace{.34\textwidth}}\\
\includegraphics[width=\textwidth]{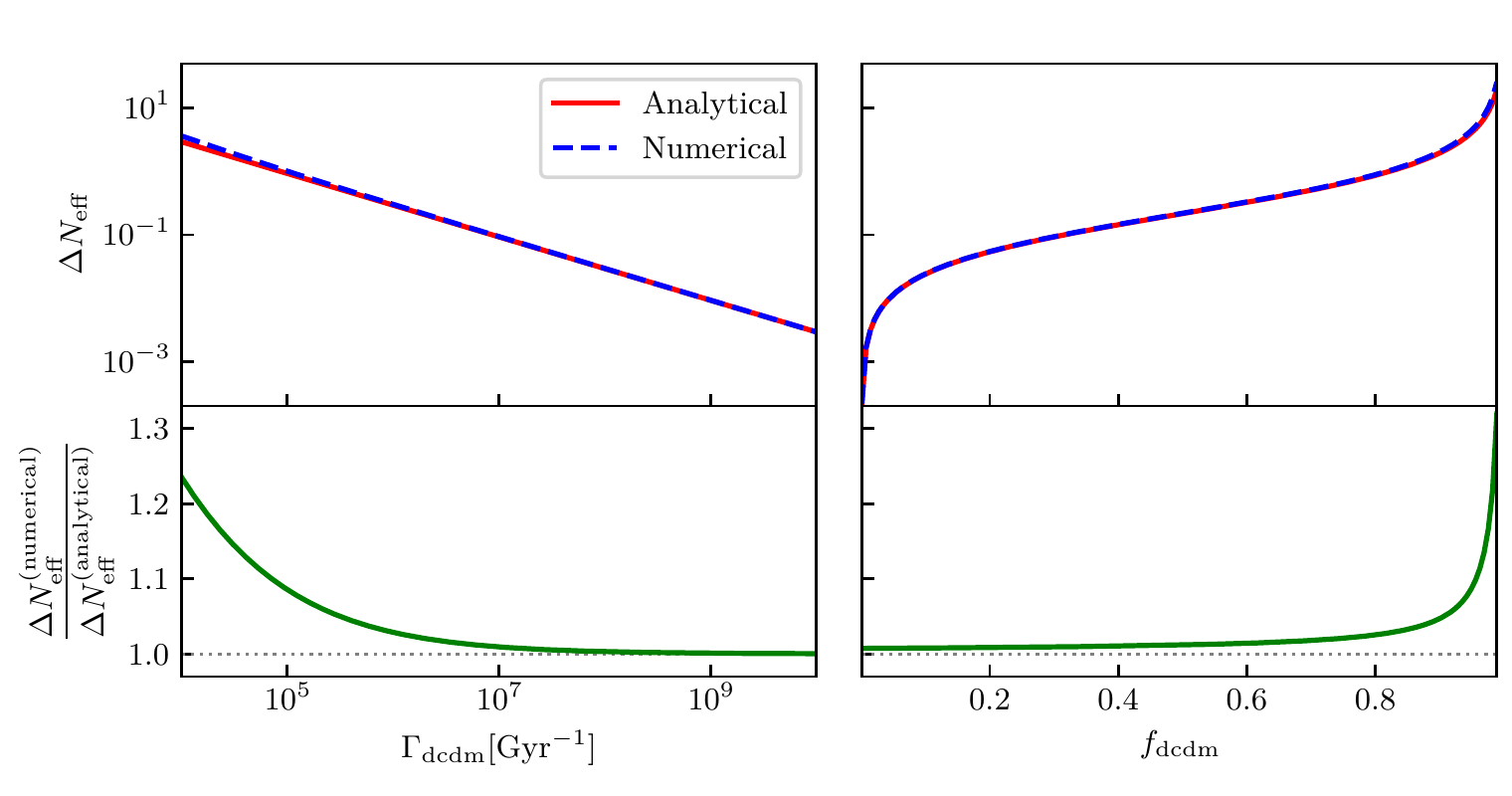} 
\caption{$\Delta N_{\rm eff}$ as a function of different parameters for both the numerical result calculated using eq.~\eqref{eq:dN_eff_class} and the analytical result calculated using eq.~\eqref{eq:neff2}. The lower panels show the ratio between the numerical and analytical $\Delta N_{\rm eff}$ as a function of the same parameters. The additional parameters used for the simulations are found in table~\ref{tab:class_input}. (a) $\Delta N_{\rm eff}$ as a function of the decay rate, $\Gamma_{\rm dcdm}$, with the fractional amount of DCDM fixed to $f_{\rm dcdm}=0.3$. (b) $\Delta N_{\rm eff}$ as a function of the fractional amount of DCDM, $f_{\rm dcdm}$, with the decay rate fixed to $\Gamma_{\rm dcdm}=10^{7}[{\rm Gyr^{-1}}]$.\label{fig:dN_eff}}
\end{figure}

At first, we are interested in how well our analytical approach fits the numerical results. Using eq.~\eqref{eq:dN_eff_class}, we calculate $\Delta N_{\rm eff}$ for different decay rates, $\Gamma_{\rm dcdm}$, and fractional amounts of DCDM, $f_{\rm dcdm}$, and plot the results as functions hereof. This can be seen in the figures~\ref{fig:dNeff_Gamma} and \ref{fig:dNeff_f} respectively along with the analytical results and the ratio between the two. The numerical and analytical results agree with increasing precision for higher values of $\Gamma_{\rm dcdm}$, which is also supported by their ratios approaching unity. Here we also note that both results approach zero for increasing decay rate, which supports that we should recover the $\Lambda$CDM model for $\Gamma_{\rm dcdm}\to\infty$ as we argued in the beginning of section~\ref{sec:Neff_mapping}. From figure~\ref{fig:dNeff_f} we, however, see that the analytical and numerical results agree with decreasing precision for higher values of $f_{\rm dcdm}$, which makes sense according to eq.~\eqref{eq:neff2} due to the divergence at $f_{\rm dcdm}=1$. It should of course behave this way because we cannot have that short-lived DCDM comprises all of the dark matter, since that would correspond to no dark matter at all after recombination, which contradicts some assumptions made in the derivation in section~\ref{sec:Neff_analytical}. Note that the value of $\Delta N_{\rm eff}$ also approaches zero for $f_{\rm dcdm}\to 0$ which again leads to the $\Lambda$CDM model. We also note that the ratio between the analytical and numerical results in figure~\ref{fig:dNeff_f} does not approach unity exactly, but rather a slightly higher value, since the decay rate is fixed at a finite value ($\Gamma_{\rm dcdm}=10^{7}\,{\rm Gyr}^{-1}$) and the ratio only approaches unity for $\Gamma_{\rm dcdm}\to\infty$, according to figure~\ref{fig:dNeff_Gamma}.

We can now run \CLASS\, again without DCDM, but with the corresponding $\Delta N^{\rm (numerical)}_{\rm eff}$ from the DCDM computation instead. According to the theory, we would expect the cosmology of such a run to be similar to that of very short-lived DCDM. Figure~\ref{fig:CMB} shows the CMB TT power spectra of the simulations with DCDM and $\Delta N_{\rm eff}=\Delta N^{\rm (numerical)}_{\rm eff}$ for two values of the decay rate, $\Gamma_{\rm dcdm}\in\{10^5\,{\rm Gyr}^{-1},\; 10^8\,{\rm Gyr}^{-1}\}$, and a fraction of initial DCDM to all dark matter of $f_{\rm dcdm}=0.2$. From this figure it is very clear that the mapping to $\Delta N_{\rm eff}$ becomes more accurate for higher values of $\Gamma_{\rm dcdm}$, as predicted by the theory. For the lower value of $\Gamma_{\rm dcdm}=10^5\,{\rm Gyr}^{-1}$, we still see the same behaviour in the power spectra, but they do not overlap as well due to the decay happening much closer to recombination, so a more significant amount of DR production is still ongoing at the time of the primary CMB signal formation. We also note that the magnitude of the relative power spectra is decreasing for larger $\Gamma_{\rm dcdm}$ as we would also expect from the theory.

\section{Current constraints on decay parameters}\label{sec:MCMC}

In order to obtain constraints on decay parameters we have used the publicly available code \textsc{MontePython}~\cite{montepython}.
This Markov Chain Monte Carlo (MCMC) sampler uses the Metropolis-Hastings algorithm to sample the probability distribution assuming flat priors. The code is run with four or seven sample chains on a computer cluster until convergence of the chains. Our Gelman-Rubin criterion for convergence of the chains is $R-1\lesssim0.01$, which means that the largest $R-1$ value of any parameter should be around or smaller than $0.01$. The prior of the $\Gamma_{\mathrm{dcdm}}$ parameter is set with different upper and lower bounds depending on which regime we want to analyse. These bounds and the number of sample chains will be stated when necessary.

\subsection{Long-lived DCDM regime}
The long-lived regime with a decay rate in the interval $\Gamma_{\mathrm{dcdm}}\in [0,10^3]\,\mathrm{Gyr}^{-1}$ corresponds to anything in between the DCDM starting to decay around the time of recombination (and thus having fully decayed by now) and an infinite lifetime where none of the DCDM has decayed yet. The common thing here is that the decay has not finished before recombination which leads to an ongoing DR production after the emission of the CMB. The late decay will thus have an effect on e.g. the angular diameter distance and other background parameters.

\begin{figure}[t]
    \centering
    \includegraphics{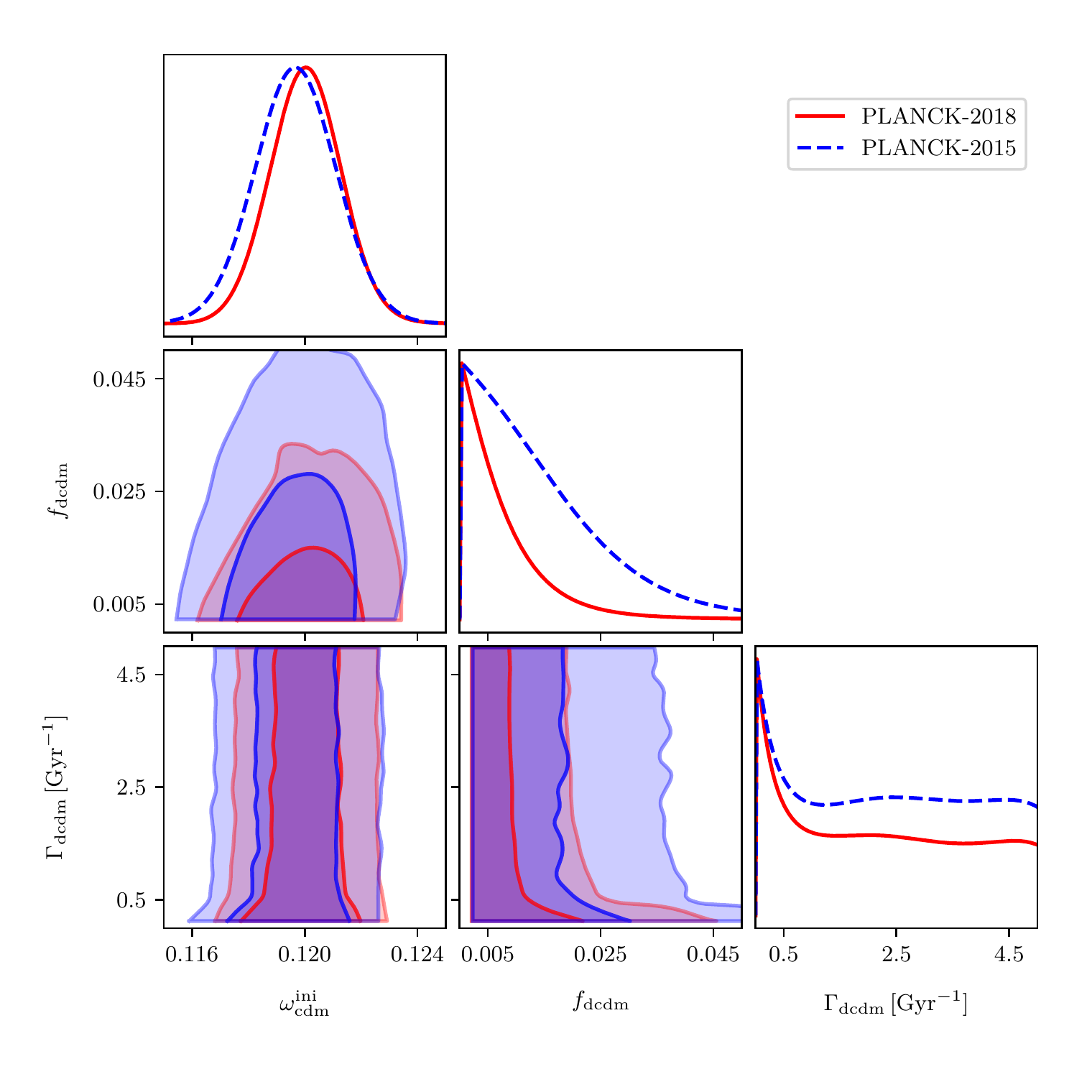}
    \caption{Triangle plot of posteriors in the long-lived regime using the two Planck data sets from 2015 and 2018.}
    \label{fig:15-18-long}
\end{figure}

\begin{figure}[t]
    \centering
    \includegraphics{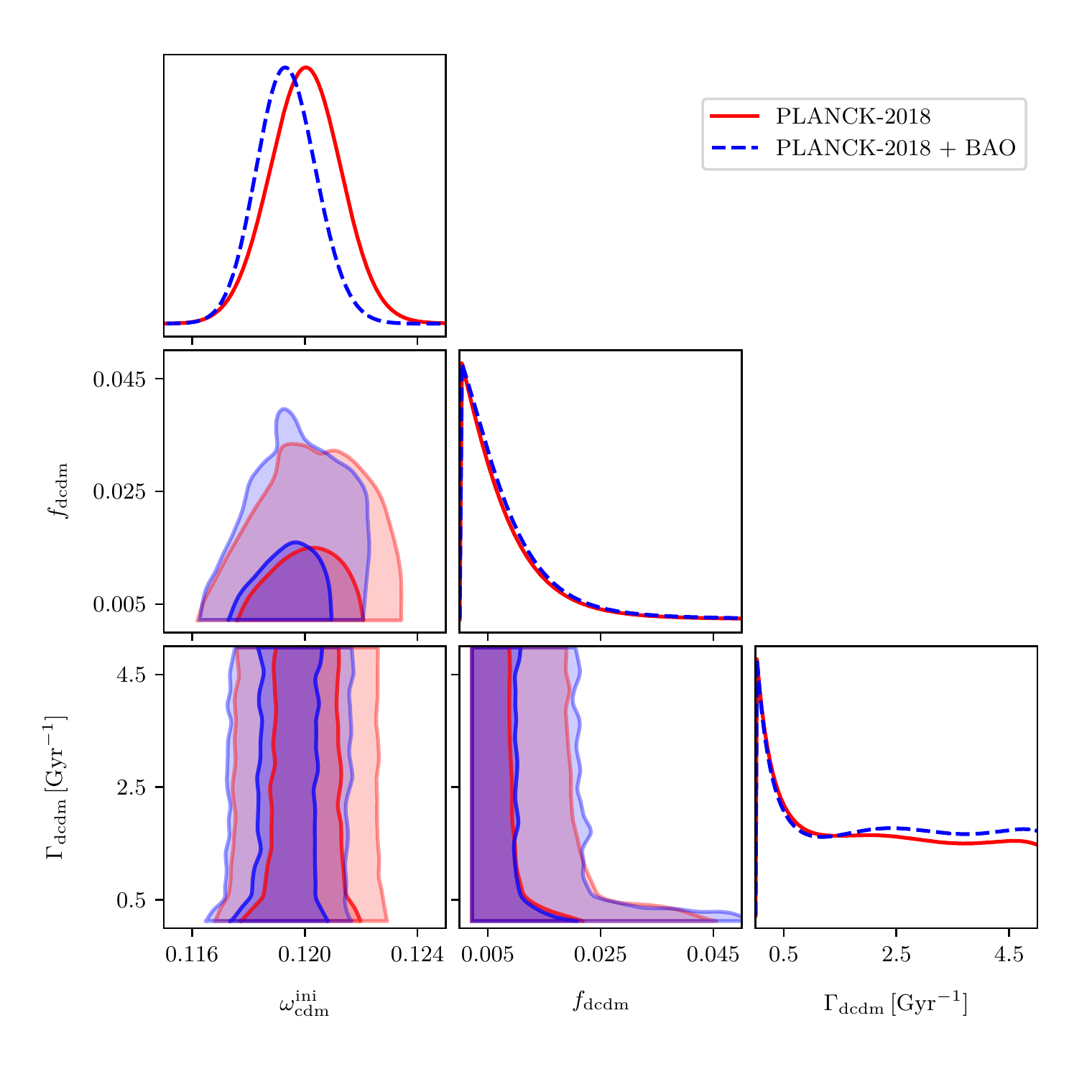}
    \caption{Triangle plot of posteriors in the long-lived regime using Planck-2018 data with and without BAO data from BOSS DR12.}
    \label{fig:BAO-long}
\end{figure}

Using \textsc{MontePython} we search the parameter space with four sample chains using the likelihoods of Planck-2015 and Planck-2018 to see the differences of the two data sets. The result of this is found in figure~\ref{fig:15-18-long}.
We have also tried to further constrain the parameters using a combination of Planck-2018 and Baryonic Acoustic Oscillation data (BAO) from BOSS DR12. These results are plotted along with that of only Planck-2018 in figure~\ref{fig:BAO-long}.

From figure~\ref{fig:15-18-long} we see that the Planck-2018 data constrains the amount of DCDM through the parameter $f_{\rm dcdm}$ much more than the Planck-2015 data does, and our results using Planck-2015 is in great agreement with those found in Ref.~\cite{Poulin_16}. The same is true regarding the constraints on the decay rate, $\Gamma_{\rm dcdm}$, which means that the Planck-2018 data favours very long-lived DCDM even more than Planck-2015 data does. The plateau for higher values of $\Gamma_{\rm dcdm}$ is due to the fact that the data cannot distinguish between the models corresponding to this plateau given that they all imply a very late decay of DCDM (much later than recombination). The plateau continues all the way to the short-lived regime, but we only show the lowest values here. The disagreement in the parameter $\omega^{\rm ini}_{\rm cdm} \equiv (\Omega^{\rm ini}_{\rm dcdm}+\Omega_{\rm cdm})h^2$ is mainly due to the disagreement in $\Omega_{\rm cdm}$ of the two Planck data sets~\cite{Planck2015,Planck2018}, since the very long-lived DCDM behaves like a stable component and we thus would not expect a significant change to the total dark matter component from that of the $\Lambda$CDM model.

In figure~\ref{fig:BAO-long} we see that the addition of BAO data relaxes the amount of DCDM through $f_{\rm dcdm}$ a little, and we see a slight change in the posterior of the decay rate as well, raising the plateau for higher values while faintly narrowing the peak towards lower values. The parameter $\omega^{\rm ini}_{\rm dcdm}$ is lowered by BAO as well, 
and this is in agreement with a narrower peak in the decay rate when including BAO, i.e. if the decay rate is smaller, the lifetime is longer, and we therefore need a smaller amount of initial dark matter since less of it has time to decay before the present.

The relevant best-fit parameters and constraints in the long-lived regime can be seen in the middle panel of table~\ref{tab:constraints} where we see a much tighter constraint on $f_{\rm dcdm}$ inferred by data from Planck-2018 as opposed to data from Planck-2015, and the same is true for the parameter $\Gamma f_{\rm dcdm}$. We also notice that the inclusion of BAO data loosens the constraints of the parameters while slightly modifying the best-fit values of $H_0$ and $\Omega^{\rm ini}_{\rm cdm}$ as well.

\subsubsection{Very long-lived DCDM regime}
In order to compare with the previous results of Ref.~\cite{Poulin_16}, we have also investigated a \textit{very long-lived} sub-regime with $\Gamma_{\mathrm{dcdm}}\lesssim H_0$. This yields slightly different results, since the long-lived runs are strongly undersampled in the very long-lived limit, but the tendencies are the same with tighter constraints from the newest Planck-2018 data. The inclusion of BAO data actually further tightens the constraint of the parameter $\Gamma f_{\rm dcdm}$ even though this was opposite in the long-lived regime.

\begin{figure}[t]
    \centering
    \includegraphics{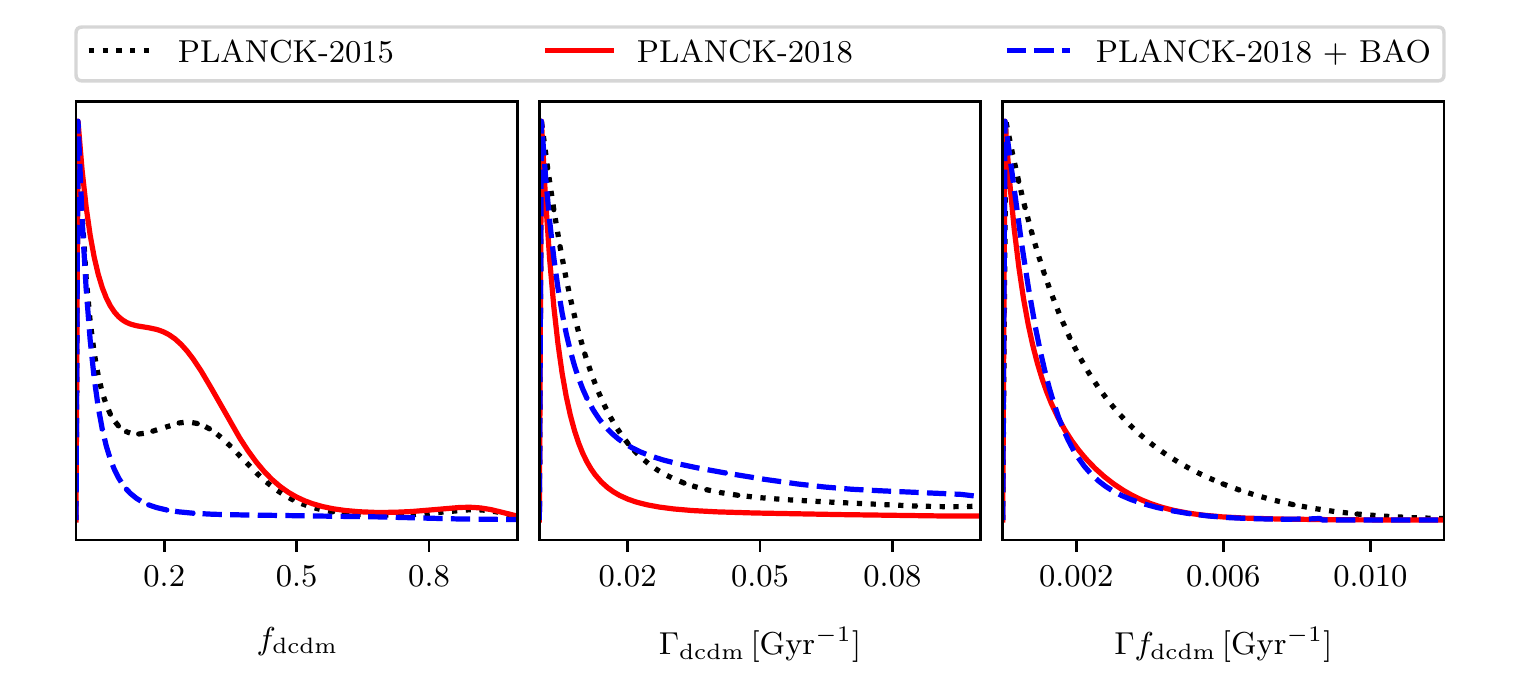}
    \caption{1D posteriors of DCDM parameters in the very long-lived sub-regime using Planck-2015 data along with Planck-2018 data with and without BAO data.}
    \label{fig:very_longlived}
\end{figure}

We find that this regime allows for a much larger fractional amount of DCDM ($f_{\rm dcdm}\to 1$), which is clear from figure~\ref{fig:very_longlived}, since the lifetime is larger than the age of the Universe and the fractional amount thus has only little impact on current observables and no impact on early-time observables such as the CMB. Constraints from this regime can be found in the upper panel of table~\ref{tab:constraints}. We notice that we are only able to constrain the parameter $f_{\rm dcdm}$ when including the BAO data, and even then it is only possible to $1\sigma$. This is due to the degenerate nature of the parameter in this regime, where all values are allowed by the data. The degeneracy is lifted when including BAO data since late-time measurements can rule out too high values of $f_{\rm dcdm}$, given that these would feel the effects of the decay as opposed to the early-time measurements. This is also apparent from figure~\ref{fig:very_longlived}.

\subsection{Short-lived DCDM regime}
In the short-lived regime, the decay rate is rather high ($\Gamma_{\mathrm{dcdm}}>10^3\,[\mathrm{Gyr}^{-1}]$) resulting in most of the DCDM decaying well before recombination. We now search the parameter space using seven sample chains, and in order to search more efficiently, we implement the logarithm of the decay rate, $\mathrm{log}_{10}(\Gamma_{\mathrm{dcdm}})$, as a parameter in \CLASS, which allows us to use this parameter in \textsc{MontePython} as well. The parameter space is effectively reduced in size which makes the sampling much faster. We cut the prior at the upper bound $\Gamma_{\mathrm{dcdm}}<10^6\,[\mathrm{Gyr}^{-1}]$ (similar to Ref.~\cite{Poulin_16}) for convergence reasons, but we have checked that the posterior for the decay rate flattens out for increasing values, thus creating another plateau where the data cannot distinguish between the models. To further increase the efficiency we use the flag \texttt{-T=2.0} (default \texttt{1.0}) when running the code, which corresponds to increasing the statistical temperature of the chains so they are more likely to jump further in the parameter space.
\begin{figure}[t]
    \centering
    \includegraphics{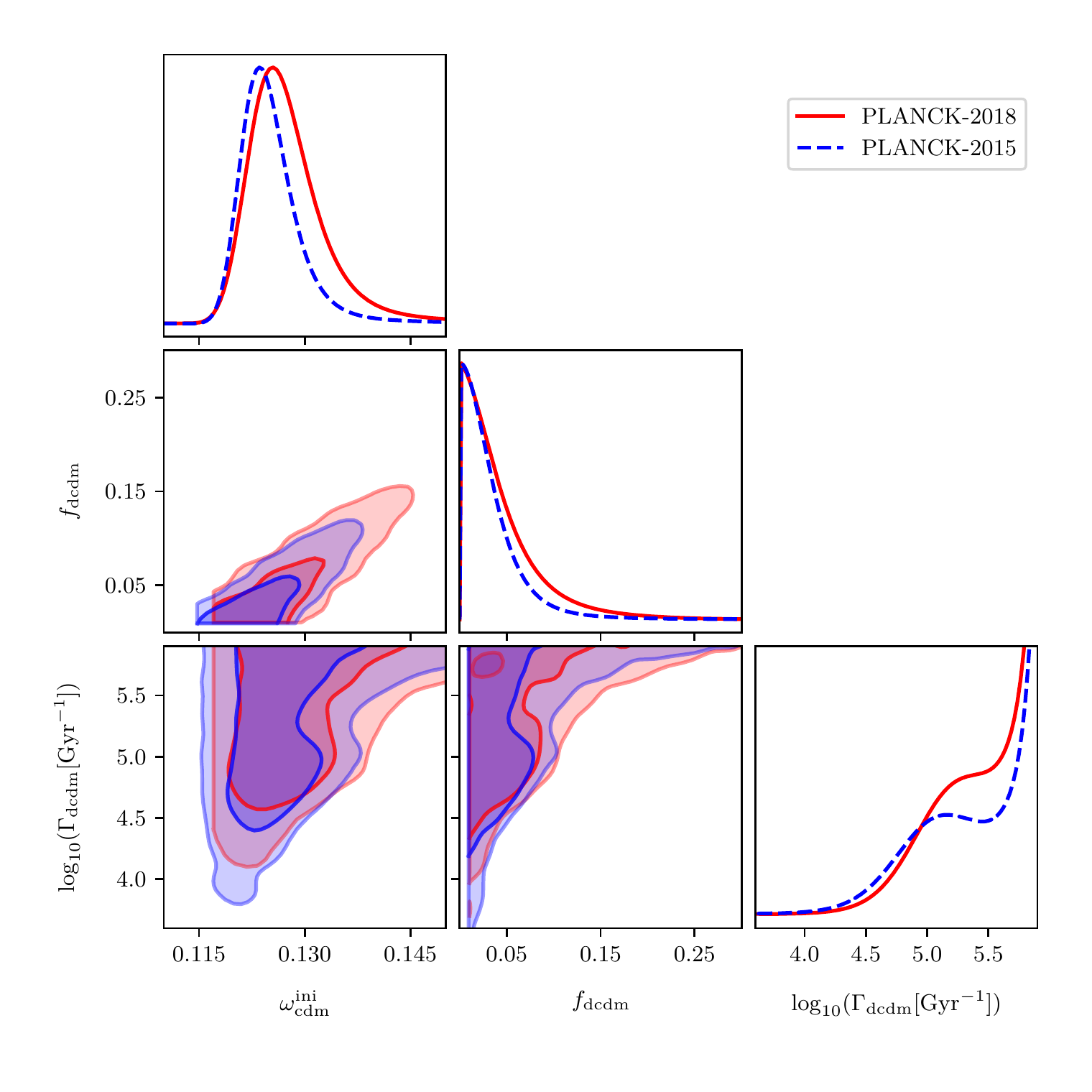}
    \caption{Triangle plot of posteriors in the short-lived regime using the two Planck data sets from 2015 and 2018.}
    \label{fig:15-18-short}
\end{figure}

\begin{figure}[t]
    \centering
    \includegraphics{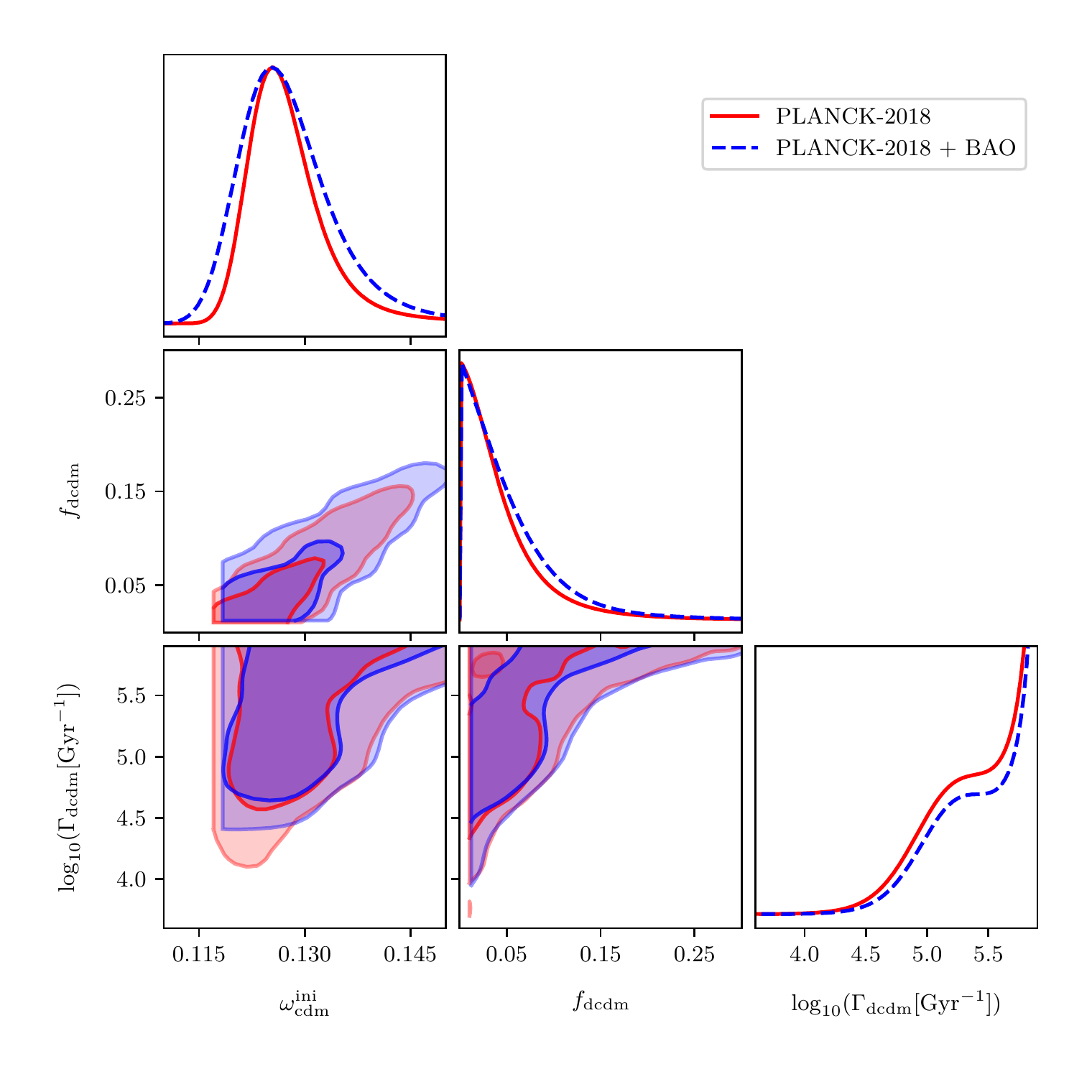}
    \caption{Triangle plot of posteriors in the short-lived regime using Planck-2018 data with and without BAO data from BOSS DR12.}
    \label{fig:BAO-short}
\end{figure}

The figures~\ref{fig:15-18-short} and \ref{fig:BAO-short} show the posterior distributions in the short-lived regime of Planck-2018 data and either Planck-2015 data or including BAO data, respectively. The 1D-posteriors of the parameter $\log_{10}(\Gamma_{\rm dcdm}[{\rm Gyr}^{-1}])$ are not shown in their entirety, but rather a closeup of the interesting region is presented by only showing the vertical interval $[0,0.2]$ (the posteriors are normalised so highest value equals unity). 

Figure~\ref{fig:15-18-short} shows that Planck-2018 data allows for a slightly higher initial dark matter density, $\omega^{\rm ini}_{\rm cdm}$, than Planck-2015 data does, which is apparent from the broader peak and longer exponential tail towards higher values in the 1D-posterior. The amount of DCDM allowed by Planck-2018 is also slightly higher which is apparent from the 1D-posterior of $f_{\rm dcdm}$ being slightly broader than that of Planck-2015. The posterior of the decay rate consists of two sub-regimes as explained in Ref.~\cite{Poulin_16}, where the small peak (plateau) in the 1D-posterior of Planck-2015 (2018) represents a decay happening mostly in between matter-radiation equality and recombination while the increase in the posterior at larger values of the decay rate represents a decay taking place mostly before matter-radiation equality. Our 1D-posterior of the decay rate for the Planck-2015 data is reminiscent of the one presented in Ref.~\cite{Poulin_16}, but with a much less significant peak even though the same data has been used. The difference here could possibly be due to the Gelman-Rubin convergence criterion of the MCMC sampling, where our criterion is an order of magnitude lower than the one used in Ref.~\cite{Poulin_16}. The conclusion, however, does not change by letting the chains converge further, but we do get that the region between the small peak and the subsequent increase is populated more frequently with a stronger convergence criterion, thus diminishing the profoundness of the peak and making it more like a plateau. Using Planck-2018 data we completely transform the peak into a plateau instead.

\begin{table}[t]
\centering
\begin{tabular}{? Sc | Sc  Sc  Sc  Sc ?}
\specialrule{.12em}{0em}{0em} 
\bf Very long-lived &
${\Omega^{\rm ini}_{\rm cdm} h^2\times 10^2}$ &
${H_0}$ &
${f_{\rm dcdm} \times 10^2}$  & $\Gamma f_{\rm dcdm}\times 10^{3}$\\
\bf regime & 
& $[{\rm km\,s^{-1}Mpc^{-1}}]$
& 
& $[{\rm Gyr^{-1}}]$\\
\hline
Planck-2015  & $11.9^{+0.3}_{-0.3}$ &  $67.7^{+1.3}_{-1.2}$  & \,\,---  & $<7.12$\\
Planck-2018  & $11.9^{+0.2}_{-0.2}$  &  $67.5^{+1.2}_{-1.2}$  & \,\,---  & $<4.01$\\
Planck-2018 + BAO  & $11.9^{+0.2}_{-0.2}$ &  $67.6^{+0.9}_{-0.9}$   & $<8.05^*$  & $<3.72$ \\
\specialrule{.12em}{0em}{0em}
\bf Long-lived regime &
${\Omega^{\rm ini}_{\rm cdm} h^2\times 10^2}$ &
${H_0}$ &
${f_{\rm dcdm} \times 10^2}$  & $\Gamma f_{\rm dcdm}\times 10^{2}$\\
& 
& $[{\rm km\,s^{-1}Mpc^{-1}}]$
& 
& $[{\rm Gyr^{-1}}]$\\
\hline
Planck-2015  & $12.0^{+0.3}_{-0.3}$ &  $67.3^{+1.6}_{-1.5}$  & $<4.14$   & $<10.99$\\
Planck-2018  & $12.1^{+0.3}_{-0.3}$  &  $67.1^{+1.2}_{-1.2}$  & $<2.44$  & $<\,\,\,5.18$\\
Planck-2018 + BAO  & $11.9^{+0.2}_{-0.2}$ &  $67.7^{+1.0}_{-0.9}$   & $<2.62$  & $<\,\,\,5.84$ \\
\specialrule{.12em}{0em}{0em} 
\bf Short-lived regime &
${\Omega^{\rm ini}_{\rm cdm} h^2\times 10^2}$ &
${H_0}$ &
${f_{\rm dcdm}}\times 10^{1}$ &  $\Gamma f_{\rm dcdm}\times 10^{-4}$\\
& 
& $[{\rm km\,s^{-1}Mpc^{-1}}]$
& 
& $[{\rm Gyr^{-1}}]$\\
\hline
Planck-2015  & $12.6^{+1.5}_{-0.5}$ &  $67.7^{+1.6}_{-1.6}$  & $<0.98$ & $<2.35$\\
Planck-2018  & $12.7^{+1.6}_{-0.8}$  &  $67.8^{+1.4}_{-1.5}$  & $<1.31$ & $<3.01$\\
Planck-2018 + BAO  & $13.1^{+1.8}_{-1.0}$ &  $68.6^{+1.2}_{-1.4}$  & $<1.49$ & $<3.78$ \\
\specialrule{.12em}{0em}{0em} 
\end{tabular}
\caption{Table of parameter constraints in the long-lived and short-lived regimes as well as in the very long-lived sub-regime. We present 2$\sigma$ constraints corresponding to a 95\% confidence level. The $*$ refers to 1$\sigma$ constraints only.}
\label{tab:constraints}
\end{table}

From figure~\ref{fig:BAO-short} it is clear that the inclusion of BAO data does not have a significant impact in the short-lived regime, as we previously argued. We see almost the exact same features in the posteriors with the exception that the 1D-posteriors of the initial dark matter density and the fractional amount of DCDM seem broader when including BAO data as opposed to just using Planck-2018 data. This is, however, not significant and we should not draw any conclusions of it, since it could very well be due to our chains not being converged enough. Increasing the amount of sample points or chains could possibly make the difference between the posteriors vanish.

The lower panel of table~\ref{tab:constraints} shows the relevant best-fit values and constraints in the short-lived regime. Here we see more loose constraints of the DCDM parameters from Planck-2018 data than from Planck-2015 data, and even more so when including BAO data. The DCDM parameter constraints are, however, not affected as much in the short-lived regime as in the very long-lived limit, where the inclusion of BAO data lifts the degeneracy of the $f_{\rm dcdm}$ parameter, and this agrees with the idea that BAO data only has little impact in short-lived regime. 

We note that at first it might seem somewhat counter-intuitive that the constraint on $f_{\rm dcdm}$ is actually loosened by the inclusion of more data, rather than strengthened. The reason for this shift is, however, easy to understand from figure~\ref{fig:BAO-short}. Given the strong correlation between $f_{\rm dcdm}$ and $\Gamma_{\rm dcdm}$, the shift towards higher values of $\Gamma_{\rm dcdm}$ enforced by the Planck-2018 (and BAO) data removes a large fraction of the low $f_{\rm dcdm}$ parameter space allowed by the Planck-2015 data. This automatically shifts the preferred range of $f_{\rm dcdm}$ upwards and leads to a less restrictive upper bound on this parameter.

\section{Impact on the Hubble and $\sigma_8$ tensions}\label{sec:tensions}

As mentioned in section~\ref{sec:previous}, the idea of decaying dark matter has been able to relieve both the Hubble tension and the $\sigma_8$ tension in matter fluctuations to some extent. 

\begin{figure}[t]
    \centering
    \includegraphics{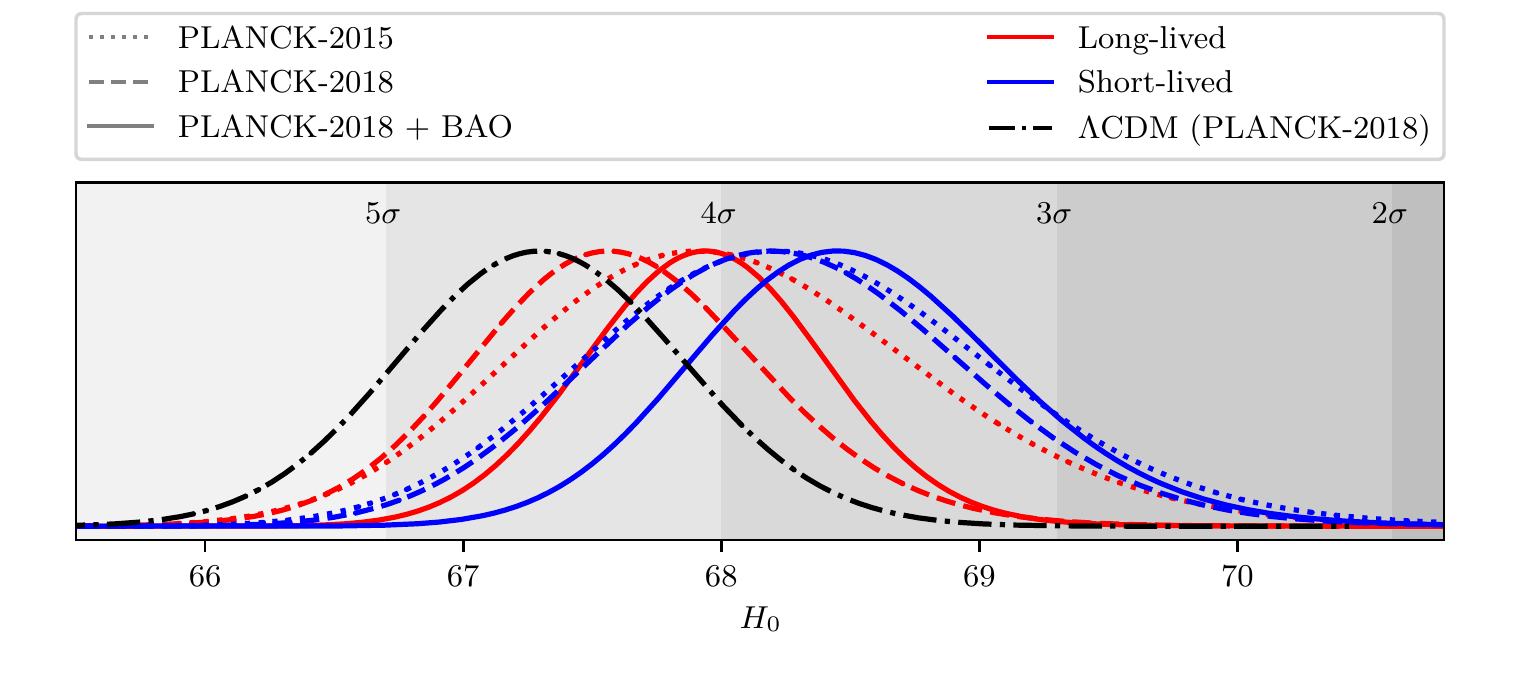}
    \caption{1D-posteriors of $H_0$ using all the data set combinations in both regimes. A posterior for the $\Lambda$CDM model using the newest Planck-2018 data has been included for reference. The shaded areas represent the latest confidence level of measurements in the local Universe of $73.2\pm1.3\,{\rm km\,s^{-1}\,Mpc^{-1}}$ from Ref.~\cite{Riess_2020}.}
    \label{fig:H0}
\end{figure}

From the 1D-posteriors of figure~\ref{fig:H0} we can clearly see that our model is indeed capable of relieving the Hubble tension to some extent using the various data sets. The best result in this regard is obtained using both Planck-2018 data and BAO data. This seems to be the case in both regimes while the short-lived regime is more successful in relieving the tension, which is also apparent from table~\ref{tab:constraints}. Compared to standard $\Lambda$CDM, our best case scenario is, however, only able to relieve the tension with $\sim\!\!1\sigma$, so the simple DCDM model does not offer the solution to this discrepancy. Perhaps one might have expected that the short-lived case would offer a better fit with high values of $H_0$. It is well known that using only CMB temperature data, there is a very strong positive correlation between $N_{\rm eff}$ and $H_0$, and since the short-lived case can be almost exactly mapped to a model with increased $N_{\rm eff}$ (see section~\ref{sec:Neff_mapping}), it is perhaps natural to expect that a high $H_0$ can be accommodated. However, the high $(H_0,N_{\rm eff})$ region is no longer allowed when polarisation data is added, thus disallowing this possibility, and in the end the preferred range of $H_0$ is only shifted marginally towards higher values in this case.

\begin{figure}[t]
    \centering
    \includegraphics{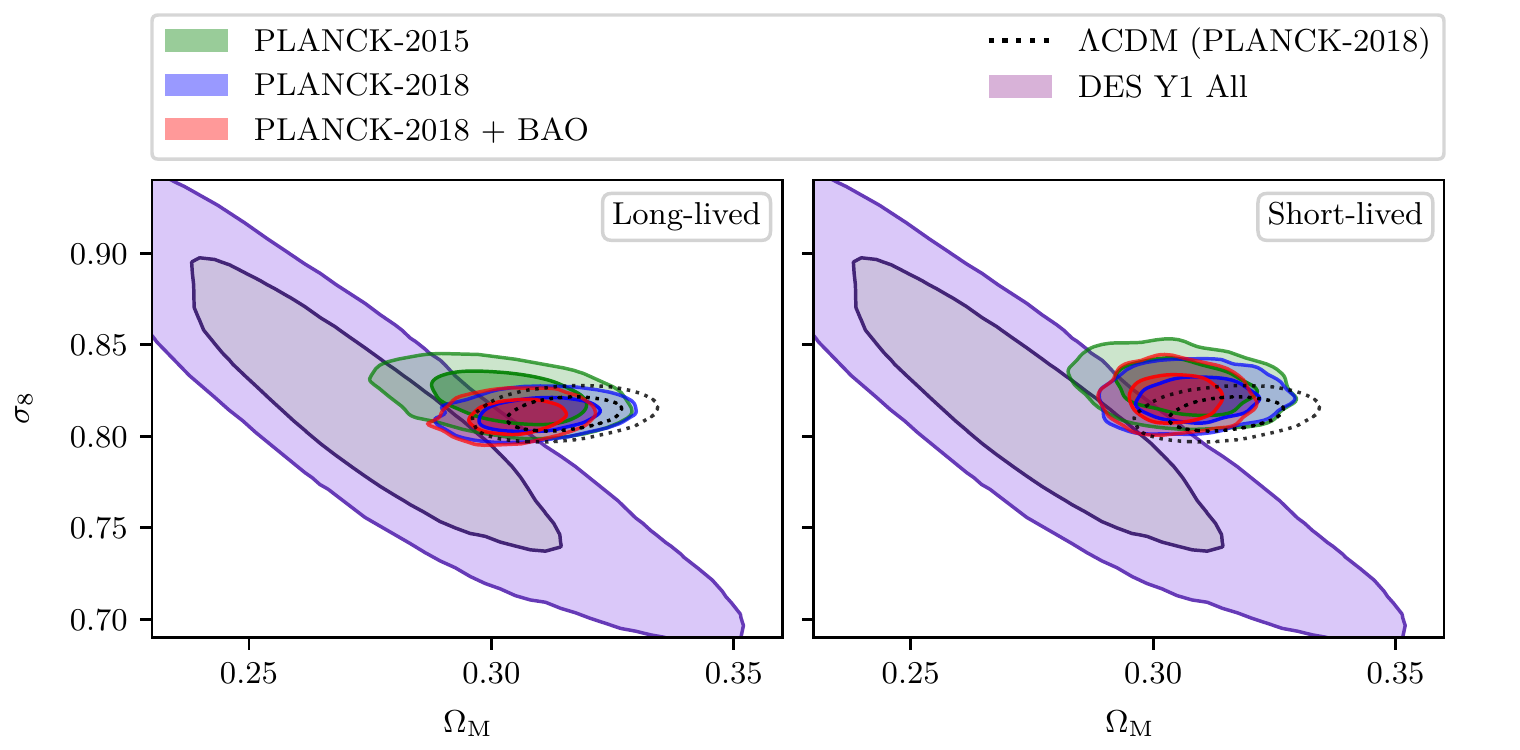}
    \caption{2D-posteriors of $\Omega_M$ and $\sigma_8$ using all the data set combinations in both regimes. A posterior from the $\Lambda$CDM model using the newest Planck-2018 data has been included for reference (dotted contours) as well as a posterior from DES Year 1 (purple contour) combining galaxy-galaxy clustering, galaxy-galaxy lensing and cosmic shear (Ref.~\cite{DES_Y1}).}
    \label{fig:sigma8}
\end{figure}

The $\sigma_8$ tension is also not solved with this model, which is apparent from the contours of figure~\ref{fig:sigma8}. This figure includes the contours of the combined DES Year 1 data from local measurements of the matter fluctuations using galaxy clustering and lensing. We can see that the contours of the long-lived regime move towards that of DES, but the tension is only relieved slightly with the 2$\sigma$ contours of the long-lived results being within the 1$\sigma$ contour of DES and vice versa. In the short-lived regime, all contours remain outside the 1$\sigma$ contour of DES, as in the case of standard $\Lambda$CDM, and they only move tangentially along the DES contour in comparison to the $\Lambda$CDM contour, thus yielding no better (nor worse) result.

Since the $\sigma_8$ tension can be relieved (slightly) only in the long-lived regime and since the Hubble tension can be relieved best in the short-lived, we recover a problem stated in Ref.~\cite{Pandey_2020} saying that attempts to relieve one of the tensions often worsens the other. We, however, do not get a worse result for the other when attempting to relieve one of the tension, but since the two tensions are relieved in different decay regimes, the simple DCDM model cannot be a solution to both tensions at once.

\section{Conclusion}\label{sec:conclusion}

Using the MCMC sampler \textsc{MontePython} and the most recent CMB data from Planck-2018, we have improved on the previous constraints on the fractional amount of DCDM, $f_{\rm dcdm}$, as well as the parameter $\Gamma f_{\rm dcdm}$ in the two regimes (long-lived and short-lived)  along with the very long-lived sub-regime. In the very long-lived sub-regime, we find that the fractional amount of DCDM cannot be constrained by Planck-2018 data alone due to the degenerate nature of the parameter in this regime. We can however get a constraint for $\Gamma f_{\rm dcdm}$ for which we find an upper bound at 2$\sigma$ of $\Gamma f_{\rm dcdm}<4.01\times10^{-3}\,{\rm Gyr^{-1}}$. In the long-lived regime, we find that the fractional amount of DCDM is much better constrained by Planck-2018 data (as opposed to Planck-2015 data) leading to an upper bound at 2$\sigma$ of $f_{\rm dcdm}<2.44\%$, and the same is true for $\Gamma f_{\rm dcdm}$ for which we find an upper bound at 2$\sigma$ of $\Gamma f_{\rm dcdm}<5.18\times10^{-2}\,{\rm Gyr^{-1}}$. The constraints in the long-lived and very long-lived regimes are thus all tighter than the constraints inferred by Planck-2015 data by a factor of $\sim\!\!2$. For the short-lived regime, the Planck-2018 data leads to more loose constraints than Planck-2015 data does, and we thus find 2$\sigma$ upper bounds of $f_{\rm dcdm}<13.1\%$ and $\Gamma f_{\rm dcdm}<3.01\times10^{4}\,{\rm Gyr^{-1}}$, which are both higher than those of Planck-2015 by a factor of $\sim\!\!1.3$.

To further constrain the decay parameters, we have included BAO data from BOSS DR-12 as well. This leads to an even tighter constraint in the very long-lived sub-regime for $\Gamma f_{\rm dcdm}$ with a 2$\sigma$ upper bound of $\Gamma f_{\rm dcdm}<3.72\times10^{-3}\,{\rm Gyr^{-1}}$, while it is now also possible to constrain $f_{\rm dcdm}$ due to the BAO data lifting the degeneracy. We then get a 1$\sigma$ upper bound of $f_{\rm dcdm}<8.05\%$. In both the long-lived and the short-lived regimes we, however, get looser constraints. In the long-lived regime we get 2$\sigma$ upper bounds of $f_{\rm dcdm}<2.62\%$ and $\Gamma f_{\rm dcdm}<5.84\times10^{-2}\,{\rm Gyr^{-1}}$, while we in the short-lived regime get $f_{\rm dcdm}<14.9\%$ and $\Gamma f_{\rm dcdm}<3.78\times10^{4}\,{\rm Gyr^{-1}}$.

The impact on the Hubble and $\sigma_8$ tensions has also been investigated by comparing our posterior distributions from the MCMC runs with the data from measurements in the local Universe, i.e. the latest value of the Hubble parameter inferred by the distance ladder in Ref.~\cite{Riess_2020} and the combined data of DES Y1 (Ref.~\cite{DES_Y1}) measuring the matter fluctuations using galaxy clustering and lensing. We get a smaller discrepancy in the Hubble parameter in the short-lived regime, reducing the tension by $\sim\!\! 1\sigma$, but in this same regime we get no improvement of the $\sigma_8$ tension. The case is opposite for the long-lived regime, where we are able to relieve the $\sigma_8$ tension slightly, but not as much in the Hubble tension. We must therefore conclude that the simple DCDM model cannot accommodate both tensions at once.

We have in addition to this investigated how the short-lived DCDM (decaying much earlier than matter-radiation equality) is analogous to a universe without DCDM but with a larger initial amount of non-EM radiation (e.g. massless neutrinos). This has been analytically mapped to a correction to the effective number of massless neutrinos species, $N_{\rm eff}$, in eq.~\eqref{eq:neff2}. Using the Boltzmann code \CLASS\, we also calculated the actual correction to $N_{\rm eff}$ and saw that this scales in the exact same way as our analytical expression. The analytical and numerical results are also shown to agree with increasing precision as $\Gamma_{\rm dcdm}\to \infty$, where we recover the standard $\Lambda$CDM model.
\\

\noindent\textbf{Reproducibility.} The modified version of \CLASS\, used to obtain the results in this paper is available at \url{https://github.com/AarhusCosmology/CLASSpp_public/} as branch 2011.01632 with SHA 767fcdec52f8135dd8cebfcba7e1b2b3cdc7bc6a. The version of \textsc{MontePython} used as well as parameter files and scripts are available at \url{https://github.com/AarhusCosmology/montepython_public/} as branch 2011.01632 with SHA e2a8af41725ce31d 63718eafe7ec614801b291f6.

\section*{Acknowledgements}
The numerical results presented in this work were obtained at the Centre for Scientific Computing, Aarhus \url{http://phys.au.dk/forskning/cscaa/}. A.N. and T.T. was supported by a research grant (29337) from VILLUM FONDEN.

\bibliographystyle{utcaps}

\bibliography{references}

\providecommand{\href}[2]{#2}\begingroup\raggedright\begin{thebibliography}{10}

\bibitem{Poulin_16}
V.~Poulin, P.~D. Serpico, and J.~Lesgourgues, ``{A fresh look at linear
  cosmological constraints on a decaying dark matter component},''
  \href{http://dx.doi.org/10.1088/1475-7516/2016/08/036}{{\em JCAP} {\bfseries
  08} (2016)  036}, \href{http://arxiv.org/abs/1606.02073}{{\ttfamily
  arXiv:1606.02073 [astro-ph.CO]}}.

\bibitem{MaBert}
C.-P. Ma and E.~Bertschinger, ``{Cosmological perturbation theory in the
  synchronous and conformal Newtonian gauges},''
  \href{http://dx.doi.org/10.1086/176550}{{\em Astrophys. J.} {\bfseries 455}
  (1995)  7--25}, \href{http://arxiv.org/abs/astro-ph/9506072}{{\ttfamily
  arXiv:astro-ph/9506072}}.

\bibitem{Ichiki_2004}
K.~Ichiki, M.~Oguri, and K.~Takahashi, ``{WMAP constraints on decaying cold
  dark matter},'' \href{http://dx.doi.org/10.1103/PhysRevLett.93.071302}{{\em
  Phys. Rev. Lett.} {\bfseries 93} (2004)  071302},
  \href{http://arxiv.org/abs/astro-ph/0403164}{{\ttfamily
  arXiv:astro-ph/0403164}}.

\bibitem{Chudaykin_2016}
A.~Chudaykin, D.~Gorbunov, and I.~Tkachev, ``{Dark matter component decaying
  after recombination: Lensing constraints with Planck data},''
  \href{http://dx.doi.org/10.1103/PhysRevD.94.023528}{{\em Phys. Rev. D}
  {\bfseries 94} (2016)  023528},
  \href{http://arxiv.org/abs/1602.08121}{{\ttfamily arXiv:1602.08121
  [astro-ph.CO]}}.

\bibitem{Berezhiani_2015}
Z.~Berezhiani, A.~Dolgov, and I.~Tkachev, ``{Reconciling Planck results with
  low redshift astronomical measurements},''
  \href{http://dx.doi.org/10.1103/PhysRevD.92.061303}{{\em Phys. Rev. D}
  {\bfseries 92} (2015) no.~6, 061303},
  \href{http://arxiv.org/abs/1505.03644}{{\ttfamily arXiv:1505.03644
  [astro-ph.CO]}}.

\bibitem{Pandey_2020}
K.~L. Pandey, T.~Karwal, and S.~Das, ``{Alleviating the $H_0$ and $\sigma_8$
  anomalies with a decaying dark matter model},''
  \href{http://dx.doi.org/10.1088/1475-7516/2020/07/026}{{\em JCAP} {\bfseries
  07} (2020)  026}, \href{http://arxiv.org/abs/1902.10636}{{\ttfamily
  arXiv:1902.10636 [astro-ph.CO]}}.

\bibitem{Chudaykin_2018}
A.~Chudaykin, D.~Gorbunov, and I.~Tkachev, ``{Dark matter component decaying
  after recombination: Sensitivity to baryon acoustic oscillation and redshift
  space distortion probes},''
  \href{http://dx.doi.org/10.1103/PhysRevD.97.083508}{{\em Phys. Rev. D}
  {\bfseries 97} (2018) no.~8, 083508},
  \href{http://arxiv.org/abs/1711.06738}{{\ttfamily arXiv:1711.06738
  [astro-ph.CO]}}.

\bibitem{Xiao_2020}
L.~Xiao, L.~Zhang, R.~An, C.~Feng, and B.~Wang, ``{Fractional Dark Matter
  decay: cosmological imprints and observational constraints},''
  \href{http://dx.doi.org/10.1088/1475-7516/2020/01/045}{{\em JCAP} {\bfseries
  01} (2020)  045}, \href{http://arxiv.org/abs/1908.02668}{{\ttfamily
  arXiv:1908.02668 [astro-ph.CO]}}.

\bibitem{Enqvist_2020}
K.~Enqvist, S.~Nadathur, T.~Sekiguchi, and T.~Takahashi, ``{Constraints on
  decaying dark matter from weak lensing and cluster counts},''
  \href{http://dx.doi.org/10.1088/1475-7516/2020/04/015}{{\em JCAP} {\bfseries
  04} (2020)  015}, \href{http://arxiv.org/abs/1906.09112}{{\ttfamily
  arXiv:1906.09112 [astro-ph.CO]}}.

\bibitem{Oldengott_2016}
I.~M. Oldengott, D.~Boriero, and D.~J. Schwarz, ``{Reionization and dark matter
  decay},'' \href{http://dx.doi.org/10.1088/1475-7516/2016/08/054}{{\em JCAP}
  {\bfseries 08} (2016)  054},
  \href{http://arxiv.org/abs/1605.03928}{{\ttfamily arXiv:1605.03928
  [astro-ph.CO]}}.

\bibitem{Blanco_2019}
C.~Blanco and D.~Hooper, ``{Constraints on Decaying Dark Matter from the
  Isotropic Gamma-Ray Background},''
  \href{http://dx.doi.org/10.1088/1475-7516/2019/03/019}{{\em JCAP} {\bfseries
  03} (2019)  019}, \href{http://arxiv.org/abs/1811.05988}{{\ttfamily
  arXiv:1811.05988 [astro-ph.HE]}}.

\bibitem{Bringmann_2018}
T.~Bringmann, F.~Kahlhoefer, K.~Schmidt-Hoberg, and P.~Walia, ``{Converting
  nonrelativistic dark matter to radiation},''
  \href{http://dx.doi.org/10.1103/PhysRevD.98.023543}{{\em Phys. Rev. D}
  {\bfseries 98} (2018) no.~2, 023543},
  \href{http://arxiv.org/abs/1803.03644}{{\ttfamily arXiv:1803.03644
  [astro-ph.CO]}}.

\bibitem{Dienes_2017}
K.~R. Dienes, F.~Huang, S.~Su, and B.~Thomas, ``{Dynamical Dark Matter from
  Strongly-Coupled Dark Sectors},''
  \href{http://dx.doi.org/10.1103/PhysRevD.95.043526}{{\em Phys. Rev. D}
  {\bfseries 95} (2017) no.~4, 043526},
  \href{http://arxiv.org/abs/1610.04112}{{\ttfamily arXiv:1610.04112
  [hep-ph]}}.

\bibitem{Raveri_2017}
M.~Raveri, W.~Hu, T.~Hoffman, and L.-T. Wang, ``{Partially Acoustic Dark Matter
  Cosmology and Cosmological Constraints},''
  \href{http://dx.doi.org/10.1103/PhysRevD.96.103501}{{\em Phys. Rev. D}
  {\bfseries 96} (2017) no.~10, 103501},
  \href{http://arxiv.org/abs/1709.04877}{{\ttfamily arXiv:1709.04877
  [astro-ph.CO]}}.

\bibitem{Vattis_2019}
K.~Vattis, S.~M. Koushiappas, and A.~Loeb, ``{Dark matter decaying in the late
  Universe can relieve the H0 tension},''
  \href{http://dx.doi.org/10.1103/PhysRevD.99.121302}{{\em Phys. Rev. D}
  {\bfseries 99} (2019) no.~12, 121302},
  \href{http://arxiv.org/abs/1903.06220}{{\ttfamily arXiv:1903.06220
  [astro-ph.CO]}}.

\bibitem{Blackadder:2014wpa}
G.~Blackadder and S.~M. Koushiappas, ``{Dark matter with two- and many-body
  decays and supernovae type Ia},''
  \href{http://dx.doi.org/10.1103/PhysRevD.90.103527}{{\em Phys. Rev. D}
  {\bfseries 90} (2014) no.~10, 103527},
  \href{http://arxiv.org/abs/1410.0683}{{\ttfamily arXiv:1410.0683
  [astro-ph.CO]}}.

\bibitem{Clark:2020miy}
S.~J. Clark, K.~Vattis, and S.~M. Koushiappas, ``{Cosmological constraints on
  late-Universe decaying dark matter as a solution to the $H_0$ tension},''
  \href{http://dx.doi.org/10.1103/PhysRevD.103.043014}{{\em Phys. Rev. D}
  {\bfseries 103} (2021) no.~4, 043014},
  \href{http://arxiv.org/abs/2006.03678}{{\ttfamily arXiv:2006.03678
  [astro-ph.CO]}}.

\bibitem{Haridasu:2020xaa}
B.~S. Haridasu and M.~Viel, ``{Late-time decaying dark matter: constraints and
  implications for the $H_0$-tension},''
  \href{http://dx.doi.org/10.1093/mnras/staa1991}{{\em Mon. Not. Roy. Astron.
  Soc.} {\bfseries 497} (2020) no.~2, 1757--1764},
  \href{http://arxiv.org/abs/2004.07709}{{\ttfamily arXiv:2004.07709
  [astro-ph.CO]}}.

\bibitem{Wang:2012eka}
M.-Y. Wang and A.~R. Zentner, ``{Effects of Unstable Dark Matter on Large-Scale
  Structure and Constraints from Future Surveys},''
  \href{http://dx.doi.org/10.1103/PhysRevD.85.043514}{{\em Phys. Rev. D}
  {\bfseries 85} (2012)  043514},
  \href{http://arxiv.org/abs/1201.2426}{{\ttfamily arXiv:1201.2426
  [astro-ph.CO]}}.

\bibitem{Abellan:2020pmw}
G.~F. Abellan, R.~Murgia, V.~Poulin, and J.~Lavalle, ``{Hints for decaying dark
  matter from $S_8$ measurements},''
  \href{http://arxiv.org/abs/2008.09615}{{\ttfamily arXiv:2008.09615
  [astro-ph.CO]}}.

\bibitem{Abellan:2021bpx}
G.~F. Abell\'an, R.~Murgia, and V.~Poulin, ``{Linear cosmological constraints
  on 2-body decaying dark matter scenarios and robustness of the resolution to
  the $S_8$ tension},'' \href{http://arxiv.org/abs/2102.12498}{{\ttfamily
  arXiv:2102.12498 [astro-ph.CO]}}.

\bibitem{Borzumati:2008zz}
F.~Borzumati, T.~Bringmann, and P.~Ullio, ``{Dark matter from late decays and
  the small-scale structure problems},''
  \href{http://dx.doi.org/10.1103/PhysRevD.77.063514}{{\em Phys. Rev. D}
  {\bfseries 77} (2008)  063514},
  \href{http://arxiv.org/abs/hep-ph/0701007}{{\ttfamily arXiv:hep-ph/0701007}}.

\bibitem{Peter:2010jy}
A.~H.~G. Peter, C.~E. Moody, and M.~Kamionkowski, ``{Dark-Matter Decays and
  Self-Gravitating Halos},''
  \href{http://dx.doi.org/10.1103/PhysRevD.81.103501}{{\em Phys. Rev. D}
  {\bfseries 81} (2010)  103501},
  \href{http://arxiv.org/abs/1003.0419}{{\ttfamily arXiv:1003.0419
  [astro-ph.CO]}}.

\bibitem{Peter:2010sz}
A.~H.~G. Peter and A.~J. Benson, ``{Dark-matter decays and Milky Way satellite
  galaxies},'' \href{http://dx.doi.org/10.1103/PhysRevD.82.123521}{{\em Phys.
  Rev. D} {\bfseries 82} (2010)  123521},
  \href{http://arxiv.org/abs/1009.1912}{{\ttfamily arXiv:1009.1912
  [astro-ph.GA]}}.

\bibitem{Wang:2014ina}
M.-Y. Wang, A.~H.~G. Peter, L.~E. Strigari, A.~R. Zentner, B.~Arant,
  S.~Garrison-Kimmel, and M.~Rocha, ``{Cosmological simulations of decaying
  dark matter: implications for small-scale structure of dark matter haloes},''
  \href{http://dx.doi.org/10.1093/mnras/stu1747}{{\em Mon. Not. Roy. Astron.
  Soc.} {\bfseries 445} (2014) no.~1, 614--629},
  \href{http://arxiv.org/abs/1406.0527}{{\ttfamily arXiv:1406.0527
  [astro-ph.CO]}}.

\bibitem{Aoyama:2011ba}
S.~Aoyama, K.~Ichiki, D.~Nitta, and N.~Sugiyama, ``{Formulation and constraints
  on decaying dark matter with finite mass daughter particles},''
  \href{http://dx.doi.org/10.1088/1475-7516/2011/09/025}{{\em JCAP} {\bfseries
  09} (2011)  025}, \href{http://arxiv.org/abs/1106.1984}{{\ttfamily
  arXiv:1106.1984 [astro-ph.CO]}}.

\bibitem{Aoyama:2014tga}
S.~Aoyama, T.~Sekiguchi, K.~Ichiki, and N.~Sugiyama, ``{Evolution of
  perturbations and cosmological constraints in decaying dark matter models
  with arbitrary decay mass products},''
  \href{http://dx.doi.org/10.1088/1475-7516/2014/07/021}{{\em JCAP} {\bfseries
  07} (2014)  021}, \href{http://arxiv.org/abs/1402.2972}{{\ttfamily
  arXiv:1402.2972 [astro-ph.CO]}}.

\bibitem{Blinov_2020}
N.~Blinov, C.~Keith, and D.~Hooper, ``{Warm Decaying Dark Matter and the Hubble
  Tension},'' \href{http://dx.doi.org/10.1088/1475-7516/2020/06/005}{{\em JCAP}
  {\bfseries 06} (2020)  005},
  \href{http://arxiv.org/abs/2004.06114}{{\ttfamily arXiv:2004.06114
  [astro-ph.CO]}}.

\bibitem{montepython}
B.~Audren, J.~Lesgourgues, K.~Benabed, and S.~Prunet, ``{Conservative
  Constraints on Early Cosmology: an illustration of the Monte Python
  cosmological parameter inference code},''
  \href{http://dx.doi.org/10.1088/1475-7516/2013/02/001}{{\em JCAP} {\bfseries
  1302} (2013)  001},
\href{http://arxiv.org/abs/1210.7183}{{\ttfamily arXiv:1210.7183
  [astro-ph.CO]}}.

\bibitem{class}
D.~Blas, J.~Lesgourgues, and T.~Tram, ``{The Cosmic Linear Anisotropy Solving
  System (CLASS) II: Approximation schemes},''
  \href{http://dx.doi.org/10.1088/1475-7516/2011/07/034}{{\em JCAP} {\bfseries
  07} (2011)  034}, \href{http://arxiv.org/abs/1104.2933}{{\ttfamily
  arXiv:1104.2933 [astro-ph.CO]}}.

\bibitem{Planck2018}
{\bfseries Planck} Collaboration, N.~Aghanim {\em et al.}, ``{Planck 2018
  results. VI. Cosmological parameters},''
  \href{http://dx.doi.org/10.1051/0004-6361/201833910}{{\em Astron. Astrophys.}
  {\bfseries 641} (2020)  A6},
  \href{http://arxiv.org/abs/1807.06209}{{\ttfamily arXiv:1807.06209
  [astro-ph.CO]}}.

\bibitem{Planck2015}
{\bfseries Planck} Collaboration, P.~Ade {\em et al.}, ``{Planck 2015 results.
  XIII. Cosmological parameters},''
  \href{http://dx.doi.org/10.1051/0004-6361/201525830}{{\em Astron. Astrophys.}
  {\bfseries 594} (2016)  A13},
  \href{http://arxiv.org/abs/1502.01589}{{\ttfamily arXiv:1502.01589
  [astro-ph.CO]}}.

\bibitem{bao}
{\bfseries BOSS} Collaboration, K.~S. Dawson {\em et al.}, ``{The Baryon
  Oscillation Spectroscopic Survey of SDSS-III},''
  \href{http://dx.doi.org/10.1088/0004-6256/145/1/10}{{\em Astron. J.}
  {\bfseries 145} (2013)  10}, \href{http://arxiv.org/abs/1208.0022}{{\ttfamily
  arXiv:1208.0022 [astro-ph.CO]}}.

\bibitem{dodelson}
S.~Dodelson, {\em {Modern Cosmology}}.
\newblock Academic Press, Amsterdam, 2003.

\bibitem{Lesgourgues:2018oca}
{\bfseries Particle Data Group} Collaboration, M.~Tanabashi {\em et al.},
  ``{Review of Particle Physics},''
  \href{http://dx.doi.org/10.1103/PhysRevD.98.030001}{{\em Phys. Rev. D}
  {\bfseries 98} (2018) no.~3, 030001}.

\bibitem{Scherrer:1984fd}
R.~J. Scherrer and M.~S. Turner, ``{Decaying Particles Do Not Heat Up the
  Universe},'' \href{http://dx.doi.org/10.1103/PhysRevD.31.681}{{\em Phys. Rev.
  D} {\bfseries 31} (1985)  681}.

\bibitem{Riess_2020}
A.~G. Riess, S.~Casertano, W.~Yuan, J.~B. Bowers, L.~Macri, J.~C. Zinn, and
  D.~Scolnic, ``{Cosmic Distances Calibrated to 1\% Precision with Gaia EDR3
  Parallaxes and Hubble Space Telescope Photometry of 75 Milky Way Cepheids
  Confirm Tension with $\Lambda$CDM},''
  \href{http://dx.doi.org/10.3847/2041-8213/abdbaf}{{\em Astrophys. J. Lett.}
  {\bfseries 908} (2021) no.~1, L6},
  \href{http://arxiv.org/abs/2012.08534}{{\ttfamily arXiv:2012.08534
  [astro-ph.CO]}}.

\bibitem{DES_Y1}
{\bfseries DES} Collaboration, T.~M.~C. Abbott {\em et al.}, ``{Dark Energy
  Survey year 1 results: Cosmological constraints from galaxy clustering and
  weak lensing},'' \href{http://dx.doi.org/10.1103/PhysRevD.98.043526}{{\em
  Phys. Rev. D} {\bfseries 98} (2018) no.~4, 043526},
  \href{http://arxiv.org/abs/1708.01530}{{\ttfamily arXiv:1708.01530
  [astro-ph.CO]}}.

\end{thebibliography}\endgroup

\end{document}